\documentclass[useAMS,usenatbib]{mn2e}
\usepackage{subfig}
\usepackage{graphicx}
\usepackage{rotating}
\usepackage{float}
\usepackage{amsmath}
\newcommand{\Mpc}{$h^{-1}$ Mpc}
\newcommand{\Msun}{$h^{-1}$ M$_{\odot}$}
\newcommand{\Mstar}{M$_{\star}$}
\newcommand{\Zsun}{Z$_{\odot}$}

\newcommand{\Mvir}{$\mathrm{M}_{\mathrm{vir}}$}
\newcommand{\Vdisp}{$\sigma$}
\newcommand{\Rvir}{$\mathrm{R}_{\mathrm{vir}}$}
\newcommand{\Mub}{M$_{\mathrm{unbound}}$}

\newcommand{\logten}{$\mathrm{log}_{10}$}
\newcommand{\La}{$\Lambda$}

\newcommand{\rw}{r$_{_{\mathrm{W}}}$}
\newcommand{\ra}{r$_{_{\mathrm{A}}}$}
\newcommand{\tH}{$t_{\mathrm{H}}$}
\newcommand{\env}{environment}

\begin{document}

	\title[Environment Redefined]{Beyond the Halo: Redefining environment with unbound matter in N-body simulations}
	\author[G. M. Shattow and D. J. Croton]{Genevieve M. Shattow \& Darren J. Croton\\
		Swinburne University of Technology, Hawthorn, VIC 3122, Australia}
	
	\maketitle
	
	\begin{abstract}
		Approximately half of the matter in the Universe is ``unbound" at $z = 0$, according to N-body simulations such as the Millennium Run. Here, we use the milli-Millennium simulation to examine the distribution of unbound matter in relation to the dark matter halos which host galaxies. We measure the unbound matter within two types of windows, using a halo dependent radius and a fixed radius at several different scales. We also consider the timescales over which a halo can accrete the local unbound matter at $z = 2$ and $z = 0$. Finally, we compare the unbound matter to observable properties of galaxies, such as local galaxy count \env\ and stellar mass. We find that halos at $z = 2$ can accrete far more of the nearby unbound matter over a Hubble time than halos at $z = 0$ and that 78\% of particles within 5 \Rvir\ of a halo at $z = 2$ will be accreted by $z = 0$, compared to 36\% of particles within 5 \Mpc\ of the halo. We also find that galaxy count \env\ is closely related to the amount of nearby unbound matter when measured on the same scale.
	\end{abstract}


\section{Introduction}
\label{unbound:intro}

Galaxies are bright and easy to observe, but on their own they are not a complete picture of the Universe. They live in the densest peaks in the cosmic web and are surrounded and connected by diffuse matter. Estimates based on hydrodynamic simulations predict that equal numbers of baryons at $z = 0$ are found in galaxies, warm-hot intergalactic matter, and cold filamentary gas \citep{Dave1999}. This diffuse gas has been observed at $z \sim 2-3$ in Ly-$\alpha$\ emission \citep{Cantalupo2014, Martin2014}, at numerous redshifts in Ly-$\alpha$\ absorption \citep{Lynds1971}, as well as in X-rays \citep{Werner2008, Ma2009, Planck2013a} at $z = 0 - 0.5$. Radio observations will be possible with future telescopes such as the {\it Square Kilometre Array} \citep{Takeuchi2014}.

The halos in which galaxies live are part of the cosmic web, a structure that has formed over the age of the universe. Structure formation in a \La CDM universe is well defined both analytically \citep{Press1974} and through N-body simulations \citep[e.g.][]{Springel2005a}. Halo growth is assumed to happen through mergers and accretion events \citep[e.g.][]{Lacey1993,Wechsler2002}, with the division between the two being somewhat arbitrary.

``Major mergers'' are usually defined as having a mass ratio between the progenitor galaxies of $M_1/M_2 < 4$ \citep[e.g.][]{Lotz2011} although sometimes the mass ratio cutoff can be as large as 10 \citep[e.g.][]{Mihos1994}. Mass ratios greater than this cutoff are considered ``minor mergers'', although sometimes these have an upper mass ratio limit of ~10. For every decade increase in mass ratio, the number of mergers increases by a factor of ~5, although this is limited to only a few decades by the mass resolution of the N-body simulation used \citep{Fakhouri2008}.

Very small mergers, or accretion evens, happen much more frequently than major mergers and contribute a significant amount of mass to an evolving halo. Unresolved mergers can account for ~40\% of the total bound mass in a halo \citep{Genel2010} although their relative importance in comparison to major and minor mergers is somewhat halo mass dependent \citep{Fakhouri2010}. Likewise, \citet{Wang2011} find that the fraction of nearby substructure that is bound to a halo depends mostly on halo mass, although the fraction of nearby substructure that is unbound to a halo is dependent on the local tidal field. Structure outside the extent of the halo is also necessary to accurately measure the matter power spectrum \citep{vanDaalen2015}.

Gas traces these additions, which can be very smooth \citep{Angulo2010}, either undergoing a shock and heating up to the virial temperature, or radiating most of its energy away and accreting as a cold flow \citep{Birnboim2003, Keres2005}.

As \citet{Fakhouri2008} discuss, N-body simulations limit the mass ratios that can be explored by the resolution of the simulation. Halos still grow through accreting the unbound particles individually or as sub-resolution halos. Gas traces these additions, either undergoing a shock and heating up to the virial temperature, or radiating most of its energy away and accreting as a cold flow \citep{Birnboim2003,Keres2005}.

It is this pre-accretion unbound matter we are most interested in. A common question addressed by galaxy environment studies is how does a galaxy's environment affect its evolution. We turn this question around to study the opposite: how does a galaxy's evolution affects its \env. One method of measuring \env, which has been used extensively in the literature, is to count the number of nearby galaxies within an aperture of fixed size. In this paper, we will continue to use this definition and explore its connection to the amount of unbound matter nearby. While both definitions can be classified as a galaxy's \env, we will use the word ``\env" to refer to the galaxy count within a fixed aperture. We will focus on three questions regarding this unbound \env:

\begin{enumerate}
	\item How is unbound matter distributed in the universe? And how does that change with time?
	\item How likely is the nearby unbound matter to accrete onto a galaxy?
	\item Are there correlations between the amount of unbound matter around a galaxy and observable properties of that galaxy that we can use to make predictions?
\end{enumerate}

This paper is organised in the following way: in Section \ref{unbound:methods}, we discuss the simulation and model used. We break our analysis into three parts to address each of the questions above. In Section \ref{unbound:unbound}, we discuss the first question of the distribution of matter in the universe and its evolution with time. In Section \ref{unbound:freefall}, we look at the free-fall times of the unbound matter and determine how likely it is that matter will accrete onto a halo. In Section \ref{unbound:observables}, we consider observable properties of galaxies and determine if there is a ``smoking gun" of future potential accretion. In Section \ref{unbound:discussion}, we extend our discussion of the three questions posed above. Finally, in Section \ref{unbound:summary}, we summarise our findings. To avoid confusion between the gas in the intergalactic medium (IGM) and the dark matter it traces, we use the term “diffuse” to refer to the former and “unbound” to refer to the latter. We also use the term IGM to refer to all gas that is not bound to a halo, not differentiating between the phases as is common in hydrodynamic simulations \citep[e.g.][]{Dave2010}.

\section{Methods}
\label{unbound:methods}

Halos, by construction, are definable objects with a finite extent in dark matter simulations. Unbound matter, however, is a nebulous term that can refer to anything from sub-resolution halos to the dark matter traced by the intergalactic medium. For the purpose of this paper, sub-resolution halos and IGM-traced structure are considered to be the same. The unbound matter, therefore, consists of the dark matter particles that are not currently part of a halo. Likewise, halos are easy to compare because they have specific properties, such as mass and extent. Unbound matter, however, exists between the halos and, beyond particle mass, does not have any properties to call its own.



\subsection{The Simulation}
\label{unbound:sim}

For our N-body simulation, we use the milli-Millennium Run (mMR), a test case for the full Millennium Simulation \citep{Springel2005a}. The mMR is a 62.5 \Mpc\ comoving periodic box with $(270)^3$ particles of mass $8.6 \times 10^8$ \Msun. It was run with cosmological parameters from the WMAP1 survey \citep{Spergel2003}. The simulation tracks perturbations from $z = 127$ until $z = 0$ over 64 snapshots. We focus on the 32 snapshots from $z = 2$ until $z = 0$, looking mostly at these two redshifts specifically. The mMR is ideal for this work because, unlike in the full Millennium, the particle data is publicly available.

We use the online database of halos and unbound particles, as determined by the halo finder SUBFIND \citep{Springel2001}. SUBFIND is a friends-of-friends algorithm that requires a certain number of particles to be linked together for a halo to be considered bound. In this paper, we use the standard minimum number of 20 particles (giving a minimum halo mass of $1.7\times 10^{10}$ \Msun) and a linking length of b = 0.2. According to SUBFIND, 23\% of the particles are considered bound at $z = 2$, and 49\% at $z = 0$. Using AHF \citep{Knollmann2009}, an adaptive mesh halo finder, 38\% of the particles are bound at $z = 0$. With Rockstar \citep{Behroozi2013}, a friends-of-friends algorithm with an adaptive linking length, 55\% of the particles are bound by $z = 0$. When using 100 particles as the minimum halo mass at $z = 0$, the same three halo finders show 41\%, 34\%, and 48\% for SUBFIND, AHF, and Rockstar, respectively.

In other tests of N-body simulations with varying box and particle size, we find the fraction of bound particles to be far more dependent on the halo finder and chosen parameters than on the simulation resolution, regardless of the resolution. This is due to the shape of the halo mass function, which when integrated between $1 \times$ M$_{\mathrm{particle}}$ and $20 \times$ M$_{\mathrm{particle}}$ (i.e. the unbound portion of the distribution) is typically 40-60\% of the total mass in the box at $z =0$ (at least for the resolution ranges cosmological simulations are current run at). It is a technical distinction and the reason why we separate the ``unbound mass" (a simulation term) and ``diffuse mass" (more physical). In the real Universe much of the diffuse mass is probably bound to their local structures out to scales larger than \Rvir. But in the simulations, as used by the semi-analytic models, this mass is thrown away.

Halos are discovered as groupings of more than 20 particles, but are re-defined in the following way. The virial mass (\Mvir) and the virial radius (\Rvir) are related by 

\begin{equation}
	\mathrm{M}_{\mathrm{vir}} = \frac{100}{G}\mathrm{H}^2(z)\ \mathrm{R}_\mathrm{vir}^3
	\label{eqn:Mvir}
\end{equation}
where G is the gravitational constant and H($z$) is the Hubble parameter \citep{Croton2006}. \Rvir\ is the radius beyond which the halo's density falls below 200 $\times$ the critical density of the Universe. The quoted values of \Mvir\ in this paper are calculated by measuring the spherical overdensity within the virial radius, rather than the number of bound particles, although these values are very similar for all but the smallest halos. We refer the reader to the SUBFIND documentation for more information.

\subsection{Considering the Unbound Matter}
\label{unbound:window}

Because unbound matter is, by definition, not part of a larger object, we consider unbound dark matter particles in relation to the halos they are near. Around each halo we define a window of radius \rw\ and count the number of unbound particles that fall within that window. If a particle falls within more than one window, it is counted as being in each window. We do this because we are interested in the bulk effect of the surrounding unbound matter and its relation to other factors, which are discussed in Section 5. We also consider doubly counted particles in Section \ref{unbound:freefall}, but individual particles are less important for the questions considered here than the net amount.

\rw\ is defined in two different ways throughout this paper. The first method is halo dependent, where we take \rw\ to be a multiple of the virial radius, \Rvir. With this metric, the windows around more massive halos will probe larger volumes. Since a halo's gravitational sphere of influence is proportional to its radius, a more massive halo influences a larger surrounding region, making this a reasonable means of defining the window. In the mMR, \Rvir\ is defined in physical coordinates and our particles' locations are in comoving coordinates. To account for this, we translate everything into comoving coordinates.

In the second method, we take \rw\ to be a fixed value, either 1, 2, or 5 \Mpc, so that each halo probes the exact same volume regardless of mass. This window is more easily comparable to observations than the halo dependent window, although it has fewer physical implications for the halo. The window radius is in comoving coordinates.

\begin{figure*} 
	\centering
	\subfloat{\includegraphics[width=\textwidth]{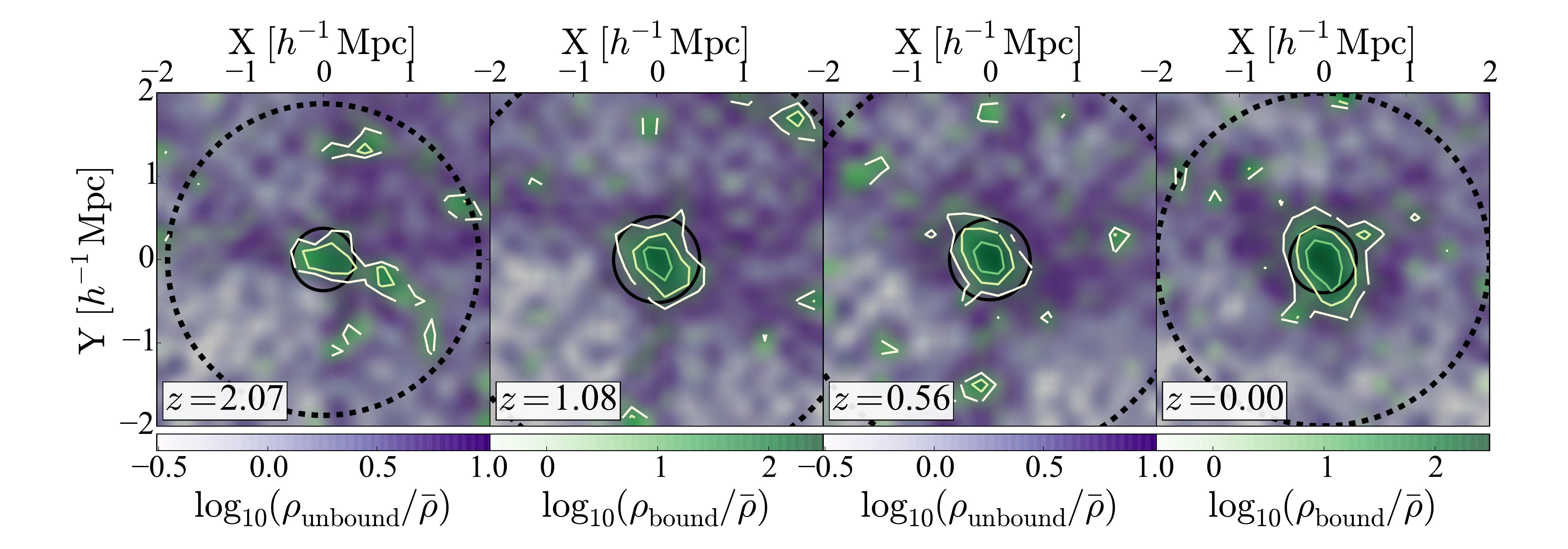}}
	\caption[Unbound mass within halo dependent windows]{The evolution of a group sized ($10^{13}$ \Msun) halo from $z = 2$ to $z = 0$. The green shading and contours represent bound matter and the purple shading represents unbound matter. Each panel is a $4 \times 4 \times 4$ \Mpc\ comoving box. The solid black circle is the virial radius of the halo, translated into comoving coordinates. The dotted black line is 5 \Rvir. \label{fig:halo_evolution}}
\end{figure*}

Figure \ref{fig:halo_evolution} shows an example of the evolution of bound and unbound matter relating to a group sized ($10^{13}$ \Msun) halo from $z = 2$ until $z = 0$ (left to right) in a $4\times4\times4$ (comoving) \Mpc\ window. The green shading and contours are the bound matter, defined by particles tagged as bound in the halo finder. The purple shading is unbound matter, defined as particles not bound to a halo. The solid black circle is 1 \Rvir, considered the extent of the halo, and the dotted circle is 5 \Rvir.

Although the bound matter extends beyond the virial radius, in our analysis we assume it fills the volume out to 1 \Rvir\ and no further. Additional bound matter can be seen in the form of smaller nearby halos. Even though they are outside the virial radius of the central halo, some of these might already be satellites.

As the halo evolves (left to right) and grows, the unbound matter appears to cluster around it, to the point where it almost forms a ring circumscribing the bound material. We will look at this evolution more closely in Section \ref{unbound:discussion}.

\subsection{The Semi-Analytic Model}
\label{unbound:SAM}

To include galaxy properties in our analysis, we post process the dark matter halos with a semi-analytic model. We use the Semi-Analytic Galaxy Evolution (SAGE) code from \cite{Croton2006}, with updates as described in Croton et al. (in prep). SAGE models the movements of baryons through reservoirs of stars, cold gas, hot gas, and ejected gas. Metals are created as part of the life cycle of stars and follow the gas, increasing the cooling. The model also includes feedback from supernovae and active galactic nuclei, which depletes the cold gas reservoir and hinders the creation of new stars.

The resolution of the semi-analytic model, when coupled with the mMR, is a stellar mass of $\sim 10^{8.5}$ \Msun. To avoid issues relating to this, we use a lower limit of stellar mass of $10^9$ \Msun. This creates a sample of 8,411 galaxies at $z = 2$ and 9,457 at $z = 0$. The full sample is our density defining population, which we use to measure the galaxy count environment. For our analysis, we further limit the sample of halos we discuss here to those containing central galaxies, eliminating satellites, and bringing the two populations to 7,089 and 6,870 for $z = 2$ and $z = 0$, respectively. Satellites are included as part of our density defining population when we discuss environment in Section 5.1, but are not included in any other capacity, as they occupy subhalos, and are therefore unlikely to accrete any additional matter. As shown in \cite{Croton2006}, SAGE agrees well with observations of the stellar mass-metallicity relation \citep[e.g][]{Tremonti2004} and the stellar mass-gas fraction relation \citep[e.g][]{Cooper2008b}, among others.

\section{Unbound Matter Around Halos}
\label{unbound:unbound}

Halos are the natural result of matter clustering, clumping, and virializing. After a halo has formed it can continue to grow through mergers and by accreting unbound matter that has clustered nearby, pulled in by gravity. When looking at the distribution of unbound matter in the universe, it is therefore natural to use halos to define the windows where we look for unbound matter. We centre a spherical window of a given size on each halo. As described above, we measure the unbound mass using two different windows, one that scales with virial radius, and one that is fixed in size.

\subsection{Halo Dependent Windows}
\label{unbound:radial}

\begin{figure*}
	\centering
	\begin{tabular}{c}
		\subfloat{\includegraphics[width=0.8\textwidth]{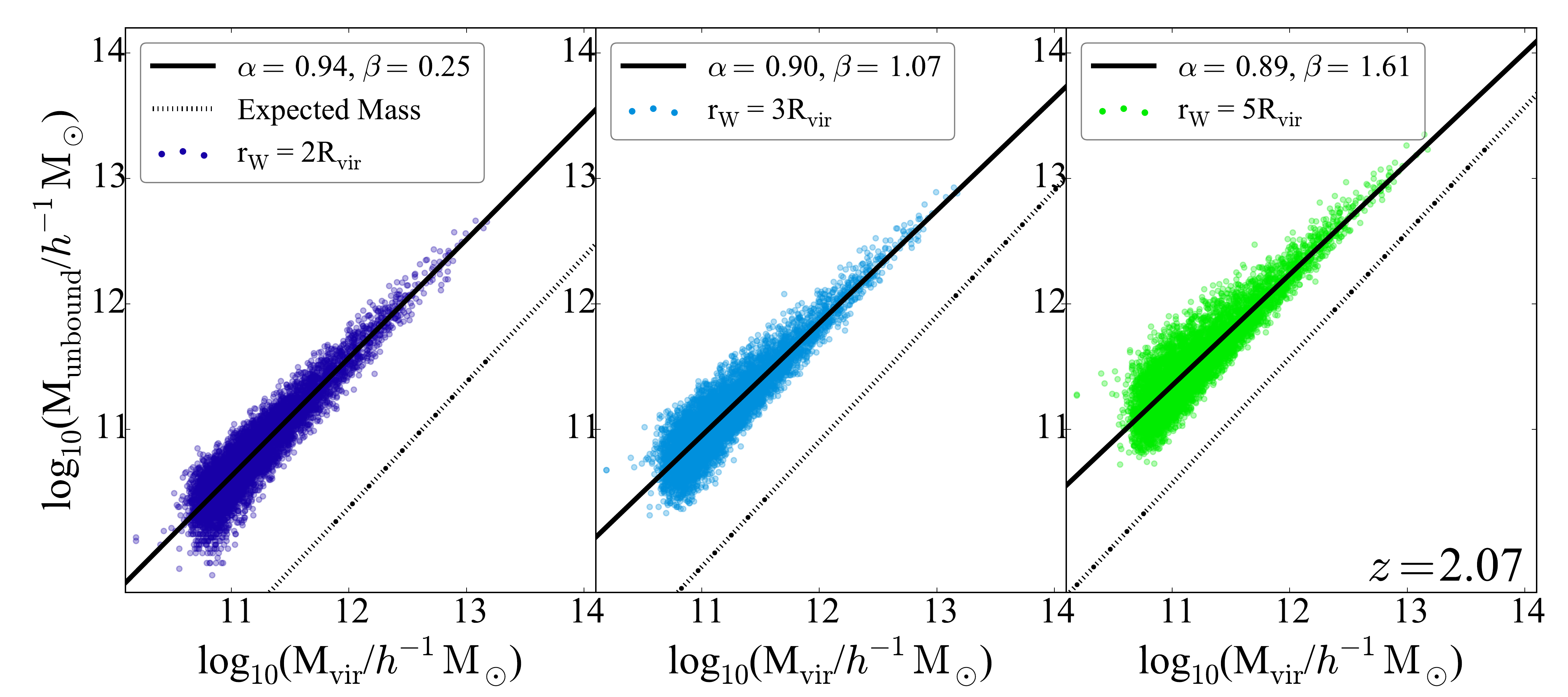}} \\
		\subfloat{\includegraphics[width=0.8\textwidth]{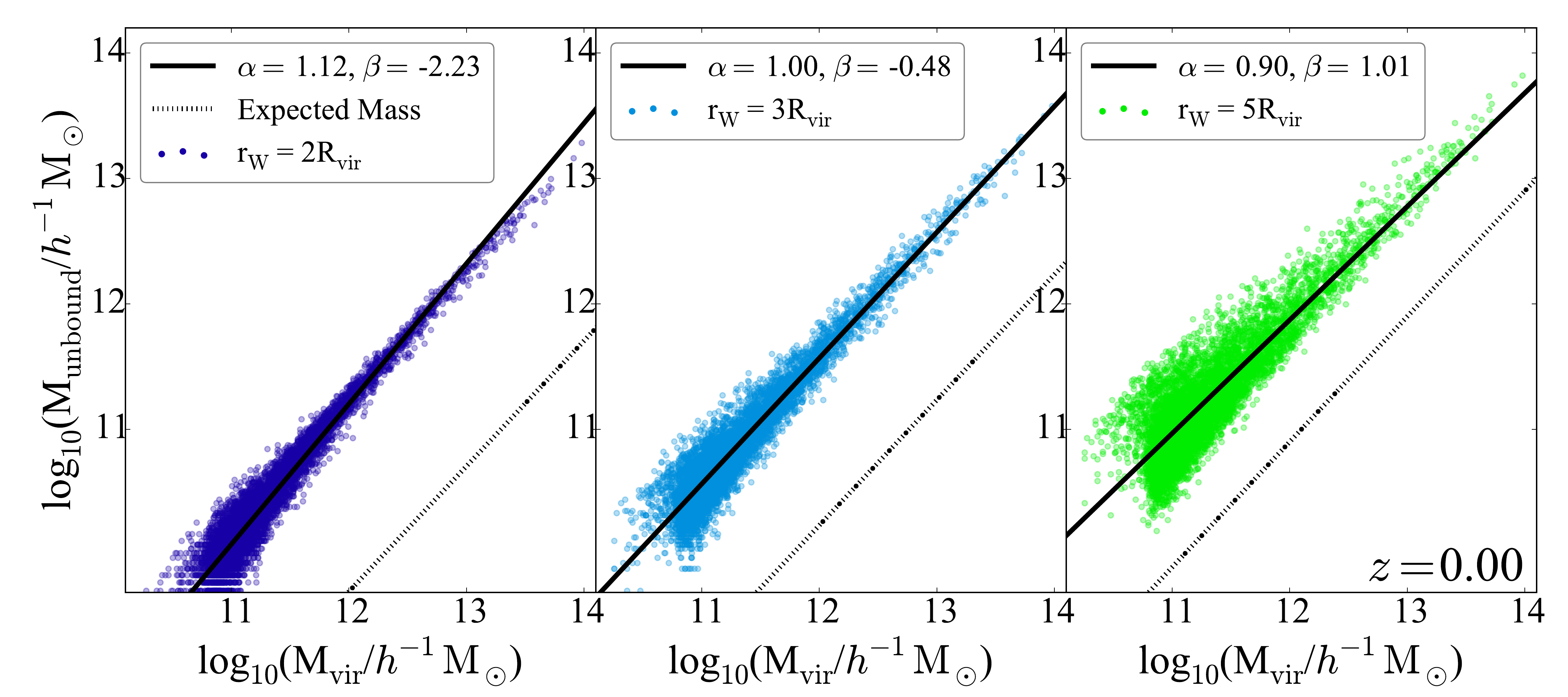}} \\
	\end{tabular}
	\caption{Halo mass versus the mass of the unbound matter contained within windows of \rw\ = 2 \Rvir\ (left), 3 \Rvir\ (middle), and 5 \Rvir\ (right) centred on each halo. The top row is $z = 2$ and the bottom row is $z = 0$. The solid black lines are a fit to the data with the parameters of the fit listed in the legend. The dotted black line is the expected mass of the unbound matter within the window if the unbound matter were evenly distributed. \label{fig:Mub_radial}}
\end{figure*}

In Figure \ref{fig:Mub_radial}, we show the virial mass of a halo against the amount of unbound mass within volumes defined by radii of 2 \Rvir, 3 \Rvir, and 5 \Rvir\ (left, middle, and right, respectively) at redshifts of $z = 2$ (top) and $z = 0$ (bottom). Each point is a halo. The solid line is a fit to the data, with $\alpha$ and $\beta$ being the slope and the intersect in the equation \logten(\Mub) = $\alpha$ \logten(\Mvir) + $\beta$. The dotted line is the mass that would be expected within the window if the unbound matter were evenly distributed throughout the box.

At both redshifts and with all three window sizes, the amount of nearby unbound matter closely tracks the virial mass of the halo. This is partially due to the window size being dependent on the central halo, although if the unbound matter were randomly distributed, there would be far less unbound matter within the window, as evidenced by the dotted line. The scatter is very low for all windows.

A slope greater than one indicates that there is more unbound matter around massive halos, relative to their size, compared to the regions around less massive halos. A slope less than one indicates that there is more unbound matter around less massive halos, relative to their size. All of our windows result in slopes very close to one, meaning there is only a slight trend with halo mass on any scale.

At both redshifts, the slope of the distribution decreases with increasing radius while the intercept increases. A decreasing slope could result from the fact that halos of all masses are likely to be within a few virial radii of massive halos, as measured by the massive halo, so although there is more matter overall near massive halos within that window, much of it is bound in other halos. The increasing intercept is due to the overall increase in the clustering of matter, despite there being fewer unbound particles at $z = 0$. The closer the fitted line (solid) is to the expected mass (dotted), the closer the unbound matter is to the average density.

From $z = 2$ to $z = 0$, the slope increases at each radius, although far more drastically (20\%) for 2 \Rvir\ than for 5 \Rvir\ (1\%). This is due to the increase in clustering of unbound matter around halos, which correlates to the increase of clustering between galaxies. The change in slope with time shows that massive halos pull in disproportionally more matter than less massive halos.

\subsection{Fixed Scale Windows}
\label{unbound:fixed}

\begin{figure*}
	\centering
	\begin{tabular}{c}
		\subfloat{\includegraphics[width=0.8\textwidth]{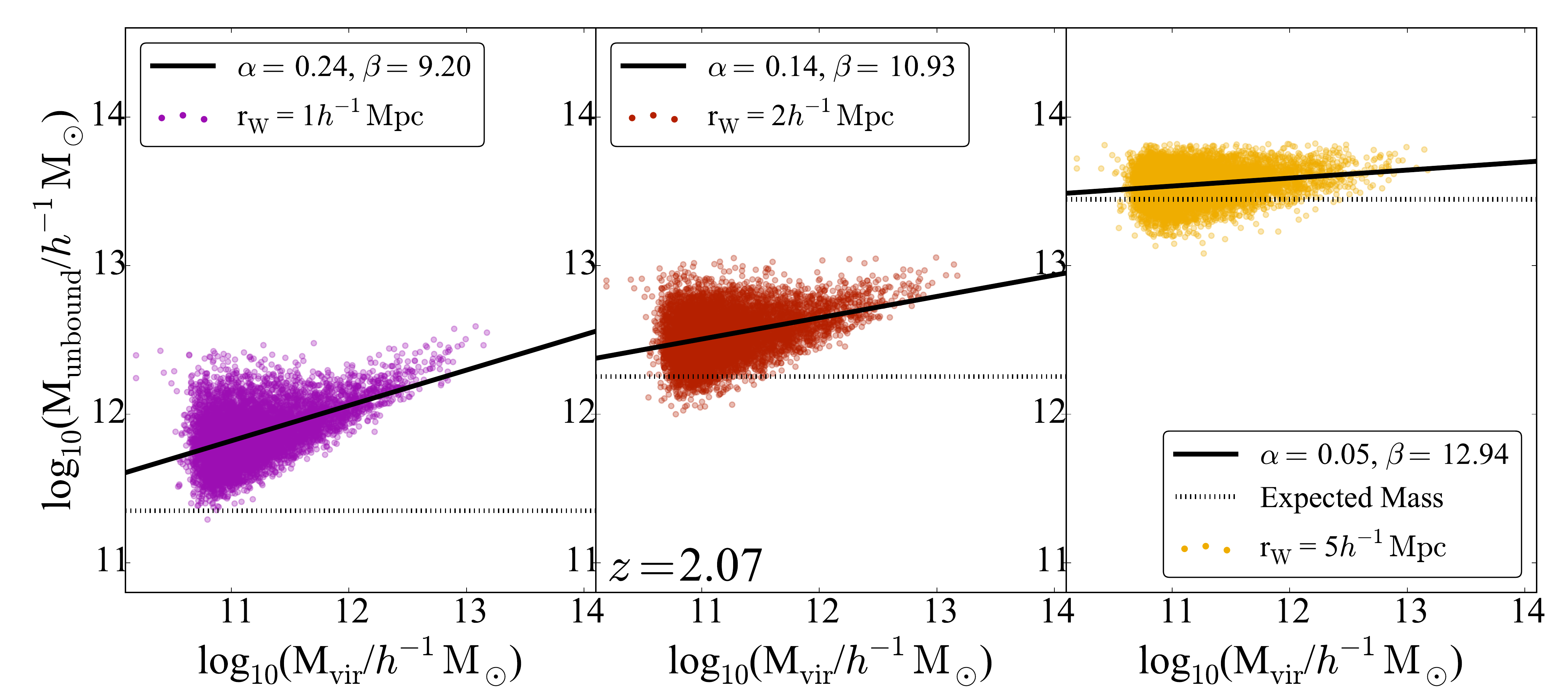}} \\
		\subfloat{\includegraphics[width=0.8\textwidth]{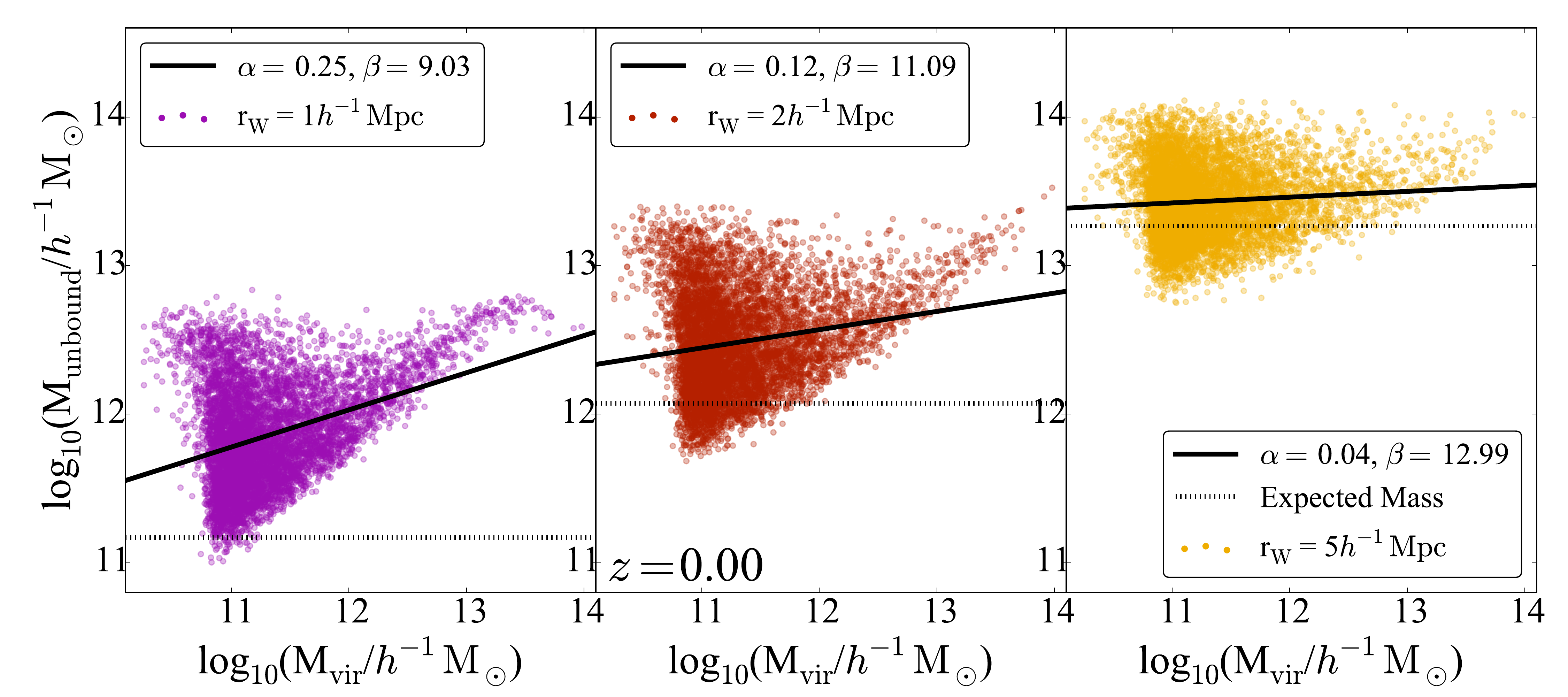}} \\
	\end{tabular}
	\caption{Halo mass versus the mass of the unbound matter in windows of 1 (left), 2 (middle), and 5 (right) \Mpc, centred on each halo. The top row is $z = 2$ and the bottom row is $z = 0$. The solid black line is a fit to the data, with the parameters of the fit being listed in the legend. The dotted line is the expected mass of the unbound matter within the window, assuming it is uniformly distributed. \label{fig:Mub_fixed}}
\end{figure*}

In Figure \ref{fig:Mub_fixed}, we again show the virial mass of the halo against the amount of nearby unbound matter, but now we are measuring the unbound matter using a window of fixed radius. This method is easier to compare to observations than a halo-based window. Left to right, \rw\ = 1, 2, and 5 \Mpc\ (purple, red, and yellow, respectively) at $z = 2$ (top) and $z = 0$ (bottom). The solid line is the fit to the data, using the parameters $\alpha$ and $\beta$ as described in Section \ref{unbound:radial}. The dotted line is the amount of unbound matter expected within the window assuming a uniform distribution. Since the window is a fixed size, the amount of unbound matter expected within it is the same, regardless of halo size.

At both redshifts and in all three window sizes, there is some dependence on halo mass. A slope of zero would signal no dependence, but each window has a positive slope, indicating that there is more unbound matter around massive halos than around less massive halos when measured using the same volume. This is related to the well known relation between halo mass and local density \citep[e.g.][]{Lemson1999, Maccio2007, Crain2009, Haas2012}. The halo mass-\env\ relation is dependent on an \env\ that is defined by the number of galaxies nearby. This figure shows that massive halos not only have more halos nearby, they also have more unbound matter.

There is a halo mass dependent lower limit to the amount of unbound matter within a fixed window, which increases with increasing halo mass. It is most striking within a 1 \Mpc\ window, although it is visible out to a radius of 5 \Mpc. There is a turn-around at \Mvir\ $\sim 10^{13}$\Msun\ in the 1 \Mpc\ window, where the amount of unbound matter within the window starts to decrease for cluster sized halos. Halos of this mass occupy a large fraction of the 1 \Mpc\ window, leaving less volume to be occupied by unbound matter.

The strongest correlation between \Mub\ and \Mvir\ comes at small scales, with the 1 \Mpc\ window having twice the slope of the 2 \Mpc\ window. The slopes are fairly stable across redshifts, as are the intercepts. The latter is an interesting result, since there is roughly half the unbound matter at $z = 0$ than at $z = 2$, as shown by the decrease in the dotted line for a given window across redshifts. It would therefore stand to reason that the intercept should decrease with time. The only scale this is true for is 1 \Mpc. For the larger scales, the slope decreases slightly and the intercept increases, also by a small amount. This stability could indicate that the unbound matter is flowing into these regions at the same rate at which it is accreting on to halos.

Unlike in the halo dependent windows, there is quite a bit of scatter in the relations between \Mvir\ and \Mub, especially at low redshift. The scatter decreases and an upward trending tail emerges for the smallest window at masses above $10^{12}$ \Msun\ at both redshifts, although it is much more pronounced at $z = 0$. In the 2 \Mpc\ window, the scatter decreases and the tail emerges at a higher mass ($\sim 10^{12.5}$ \Msun). In the largest window, the scatter decreases with increasing \Mvir, but an the tail does not emerge. A larger simulation with a higher quantity of massive halos might show the tail for the largest window, but the mMR has too few cluster sized halos.

For comparison purposes between the two window scalings, we have listed the virial radii for several masses at both redshifts of interest. They can be found in Table \ref{tab:Rvir_z}. For a $10^{12.5}$ \Msun\ halo, a 1 \Mpc\ window corresponds to a few virial radii at both redshifts of interest.

\begin{table}
	\centering
	\begin{tabular}{c c c c}
		\logten(\Mvir) & \Rvir $(z = 0)$  & \Rvir $(z = 2)$  & \Rvir $(z = 2)$  \\
		 in \Msun & in \Mpc & {\it physical} &  {\it comoving}\\
		 & & {\it coordinates} & {\it coordinates} \\
		\hline
		10.5 & 0.051 & 0.025& 0.075 \\
		11.5 & 0.111 & 0.053 & 0.159 \\
		12.5 & 0.239 & 0.115 & 0.345 \\
		13.5 & 0.514 & 0.247 & 0.741 \\
		\hline
	\end{tabular}
	\caption{The virial radii of halos at different redshifts. First column: halo mass; second column: virial radius at $z = 0$; third column: virial radius at $z = 2$ in physical coordinates; fourth column: virial radius at $z = 2$ in comoving coordinates. \label{tab:Rvir_z}}
\end{table}

\section{Free-fall Times}
\label{unbound:freefall}

Finding the amount of unbound matter around a halo will give some indication of how much that halo will grow by accretion in the future. Proximity, however, only implies that the unbound matter will fall onto the halo on a short enough timescale to make a difference in the evolution of the galaxy. To measure that timescale, we use the free-fall time of the halo plus the nearby unbound matter within our above considered windows.

The free-fall time of an object is the amount of time it takes for that object to collapse under its own gravity, assuming there are no pressure forces countering it. For a spherically symmetric object of density $\rho$, the free-fall time is 

\begin{equation}
t_{ff} = \sqrt{\frac{3 \pi}{32 \mathrm{G} \rho}},
\label{eqn:ff}
\end{equation}
where $\rho$ = M/($\frac{4}{3} \pi$ R$^3$).

For a halo in the Millennium simulation, virial mass and radius are related by Equation \ref{eqn:Mvir}, so the average density of a halo is 

\begin{equation}
\rho_{\mathrm{vir}} = \frac{\mathrm{M}_{\mathrm{vir}}}{4/3 \pi \mathrm{R}_{\mathrm{vir}}^3} = \frac{3}{4\pi} \frac{100}{\mathrm{G}} \mathrm{H}(z)^2.
\label{eqn:rhovir}
\end{equation}
Combining Equations \ref{eqn:rhovir} and \ref{eqn:ff}, the free-fall time of a halo is

\begin{equation}
t_{ff}^{\mathrm{vir}} = \frac{\pi}{20 \sqrt{2}} \mathrm{H}(z)^{-1} \simeq 0.1 t_H
\label{eqn:ffH}
\end{equation}
where $t_H$ is the Hubble time, or the age of the Universe at a given redshift, assuming expansion has been constant.

A reasonable measure of how likely a halo is to grow from accreting unbound matter can be found in the free-fall time of the halo plus the unbound matter. To calculate this, we assume spherical symmetry in the distribution of \Mub\ around the halo. While this is clearly not accurate, it is our version of assuming a spherical cow. 

We are now considering both the mass within the halo and the unbound mass out to $n\ \times$\ \Rvir, so

\begin{equation}
\mathrm{M} = \mathrm{M}_{\mathrm{vir}} + \mathrm{M}_{\mathrm{unbound}} = \mathrm{M}_{\mathrm{vir}} (1 + \mathrm{M}_{\mathrm{unbound}}/\mathrm{M}_{\mathrm{vir}}), 
\end{equation}
\begin{equation}
\mathrm{R} = n \times \mathrm{R}_{\mathrm{vir}} 
\end{equation}
Incorporating these values into $\rho$, 

\begin{equation}
\rho = \frac{\mathrm{M}_{\mathrm{vir}} (1 + \mathrm{M}_{\mathrm{unbound}}/\mathrm{M}_{\mathrm{vir}})}{(\frac{4}{3} \pi n^3 \mathrm{R}_{\mathrm{vir}}^3)} = \rho_{\mathrm{vir}} \times  \frac{1 + \mathrm{M}_{\mathrm{unbound}}/\mathrm{M}_{\mathrm{vir}}}{n^3}
\end{equation}
and the free-fall time from Equations \ref{eqn:ff} and \ref{eqn:ffH} is now

\begin{equation}
t_{ff} = t_{ff}^{\mathrm{vir}} \sqrt{\frac{n^3}{1 + \mathrm{M}_{\mathrm{unbound}}/\mathrm{M}_{\mathrm{vir}}}}.
\label{eqn:free-fall-unbound}
\end{equation}

This relation shows that the free-fall time of the unbound matter relies on both the spatial extent of the window, as well as the amount of unbound matter within that volume. At a given radius around a halo of a given mass, a larger amount of unbound matter will decrease the free-fall time.

\begin{figure*}
	\centering
	\begin{tabular}{c}
		\subfloat{\includegraphics[width=0.8\textwidth]{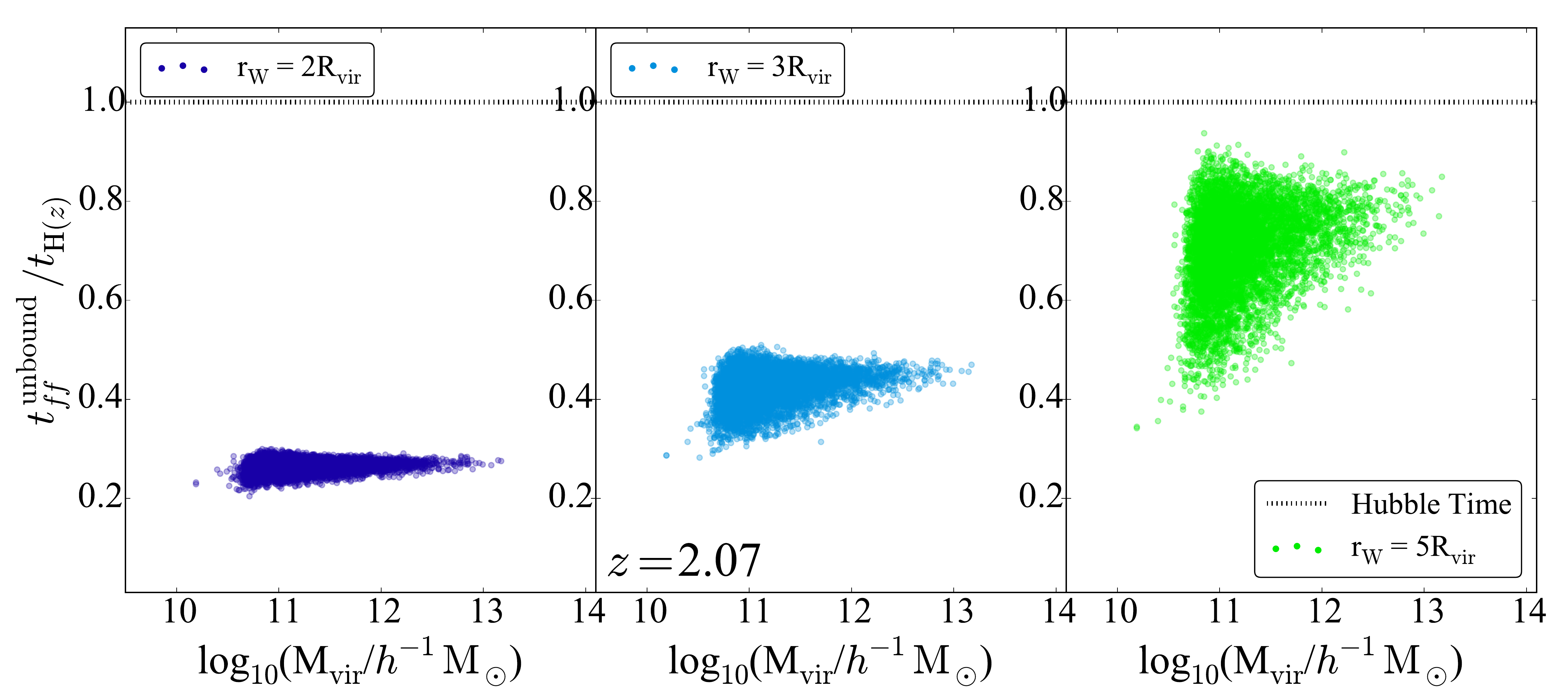}} \\
		\subfloat{\includegraphics[width=0.8\textwidth]{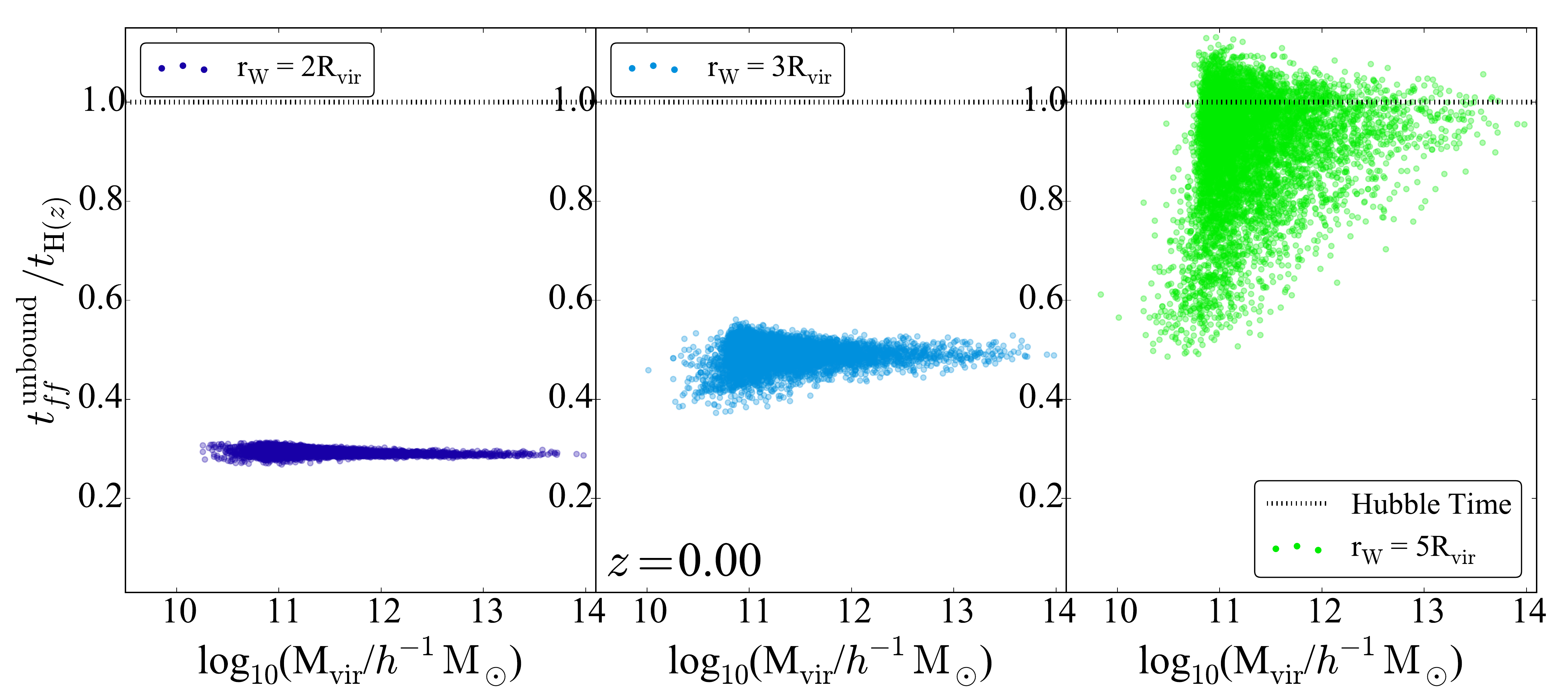}} \\
	\end{tabular}
	\caption{The halo mass versus the free fall time of the halo plus the unbound matter as measured in windows of \rw\ = 2 \Rvir\ (left), 3 \Rvir\ (middle), and 5 \Rvir\ (right) at $z = 2$ (top row) and $z = 0$ (bottom row). The dotted line is the Hubble time. \label{fig:Mub_radial_ff}}
\end{figure*}

Figure \ref{fig:Mub_radial_ff} shows the free-fall time of the unbound matter enclosed within 2 \Rvir, 3 \Rvir, and 5 \Rvir\ (left to right) at $z = 2$\ and $z = 0$\ (top and bottom, respectively) as a fraction of the Hubble time at that redshift. The dotted line in each panel is 1 Hubble time.

At $z = 2$, all of the unbound matter out to 5 \Rvir\ around all of the halos in the sample can accrete onto the galaxies within a Hubble time. At $z = 0$, the unbound matter is more clustered, meaning the distribution is steeper, so the difference in free-fall times between window sizes is more pronounced. All of the galaxies have the potential to accrete all of the nearby unbound matter out to 2 and 3 \Rvir, but only 80\% of the galaxies can accrete all of the unbound matter out to 5 \Rvir.

As in Section \ref{unbound:unbound}, we continue by comparing the halo dependent window to the more observationally motivated fixed radius window. To calculate the free-fall time of unbound matter within a fixed window, we use Equation \ref{eqn:free-fall-unbound} and define $n$\ as

\begin{equation}
n = \mathrm{r}_{\mathrm{W}}/\mathrm{R}_{\mathrm{vir}}.
\end{equation}

\begin{figure*}
	\centering
	\begin{tabular}{c}
		\subfloat{\includegraphics[width=0.8\textwidth]{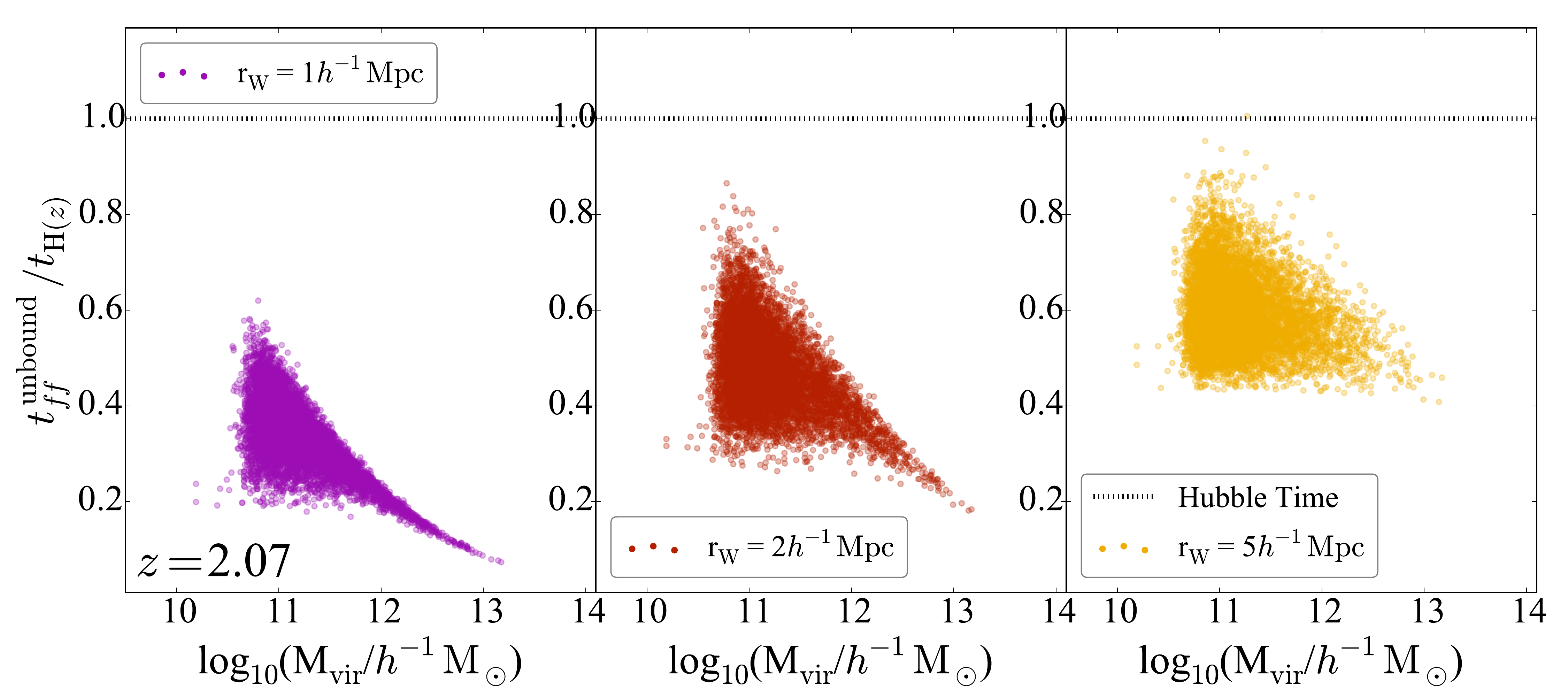}} \\
		\subfloat{\includegraphics[width=0.8\textwidth]{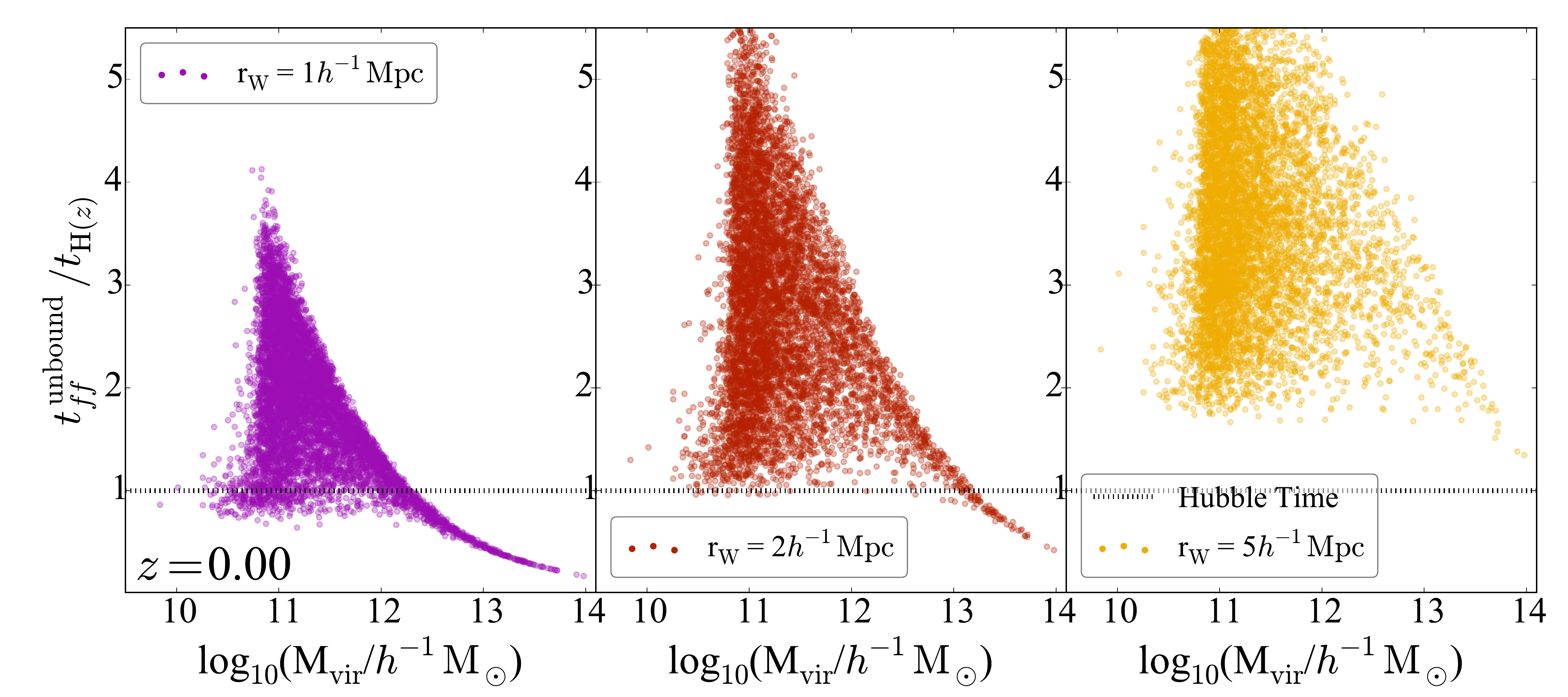}} \\
	\end{tabular}
	\caption{The halo mass versus the free fall time of the halo plus the unbound matter as measured in windows of \rw\ = 1 (left), 2 (middle), and 5 (right) \Mpc. The dotted line is the Hubble time. Note the change in range on the y-axis between $z = 2$ (top row) and $z = 0$ (bottom row). \label{fig:Mub_fixed_ff}}
\end{figure*}

Figure \ref{fig:Mub_fixed_ff}\ shows the free-fall times of the unbound matter within fixed windows of radii 1, 2, and 5 \Mpc\ (left to right) at $z = 2$ and 0 (top and bottom, respectively). The dotted line is 1 Hubble time. Note the significant change in the range of the y-axis.

At $z = 2$, all of the galaxies can accrete all of the unbound matter out to 5 \Mpc\ within a Hubble time. At $z = 0$, however, very few halos have the potential to accrete all of their nearby unbound matter within a Hubble time. Only 12\% of galaxies can accrete everything within 1 \Mpc, 1\% of halos can accrete everything within 2 \Mpc, and no galaxies will accrete everything within 5 \Mpc. Since the matter in the 5 \Mpc\ window includes the matter in the 1 and 2 \Mpc\ windows, some halos can still accrete some of the unbound matter within this volume. Our calculations reflect the potential of a halo accreting all of the unbound matter within the window.

While only a small percentage of sub-L* halos have the potential to accrete all of the unbound matter within 1 \Mpc, all of the halos over $\sim 10^{12.2}$ \Msun\ can do so. That this turn-over occurs at L* is most likely a numerical coincidence, but it is one we can potentially exploit future analysis. A fixed window of 1 \Mpc\ is a better predictor of future growth for galaxies in halos more massive than $10^{12.2}$ than it is for less massive halos. It is also a better predictor when compared to windows of larger radii.

As mentioned above, in our current model, individual unbound dark matter particles can be counted more than once if they are near more than one halo. The fraction of doubly (or more) counted particles within 5 \Rvir\ of a halo is 39\% at $z = 2$ and 46\% at $z = 0$, with each particle found in a window being counted an average of $\sim 1.7$ times. This fraction is even higher for particles within 5 \Mpc\ of a halo, with the fractions being close to 100\% at both redshifts, and each particle being counted $\sim$37 times at $z = 2$ and $\sim$70 times at $z = 0$. If each dark matter particle were counted only once based on the gravitational pull of each nearby halo, the amount of dark matter around a given halo would decrease (dramatically for matter calculated within a fixed window) and the free-fall times would increase. This would affect sub-L* halos more than massive halos, since the former have smaller gravitational potential wells and would therefore be less likely to pull the particles in.

The free-fall time only measures the potential of a halo to accrete nearby unbound matter and inherently does not include any pressure forces or angular momentum that would prevent collapse. We are not interested in the collapse of the system, but rather the accretion of the unbound dark matter particles onto the halo, so the free-fall time is a proxy for a more complex measurement which considers the time it would take for a given particle to reach the edge of a halo. We expect this to correlate with the infall of diffuse gas that could ultimately go on to fuel future star formation and galaxy growth, and hence we can speculate about the fate of the galaxy itself.

\section{Observables}
\label{unbound:observables}

The correlation between halo mass and the amount of nearby unbound matter measured in Section \ref{unbound:unbound} shows that halo mass is a reasonable predictor of nearby unbound matter, depending on window type and scale. Unfortunately, halo mass is not an easily measurable quantity observationally. This, in turn, makes \Rvir\ an unreliable basis for a metric. To make our results more comparable to observations, we now switch to using only a fixed window for measuring the unbound matter and extend the previous analysis to compare the nearby unbound matter with mock galaxy properties. From this we can make predictions of how diffuse matter is distributed which can ultimately be tested by observers using instruments, such as the Cosmic Web Imager \citep{Martin2014}, or new telescopes, such as the {\it SKA}, to measure the intergalactic gas which traces the unbound matter.

\subsection{Galaxy Environment}
\label{unbound:env}

An obvious first ``observable" to look at is the \env\ of the galaxy as measured by the nearby galaxies. We use a spherical fixed aperture metric to define \env, which has been discussed extensively in \cite{Muldrew2012}. To avoid confusion, we will continue to use the word ``window" to describe the volume containing unbound matter and use ``aperture" to describe the volume over which we are counting galaxies. It is useful to note that windows are centred on the halos and the apertures are centred on the galaxies. Within a semi-analytic model, these are the same location, but this is not always true in hydrodynamic simulations or in observations.

A galaxy or halo's local density is defined as 

\begin{equation}
	\rho/\bar{\rho} = \frac{\mathrm{N}_{\mathrm{g}}}{4/3 \pi \mathrm{r}_{\mathrm{A}}^3}\ \frac{1}{\bar{\rho}}
\end{equation}
where N$_\mathrm{g}$ is the number of galaxies within the aperture (including itself, other central galaxies, and any satellite galaxies that fall within the aperture), \ra\ is the radius of the aperture, and $\bar{\rho}$ is the mean number of density defining galaxies per unit volume.

For reference, a single galaxy in our simulation within a 1 \Mpc\ aperture has a $\rho/\bar{\rho} = 6.16$, in a 2 \Mpc\ aperture $\rho/\bar{\rho} = 0.77$, and in a 5 \Mpc\ aperture $\rho/\bar{\rho} = 0.05$. By definition, all galaxies are over-dense when measured on a 1 \Mpc\ scale. As has been discussed in \citet{Muldrew2012, Haas2012, Shattow2013}, etc., the \env\ of a galaxy, as measured by galaxy counts within an aperture, correlates with the mass of the halo containing the galaxy.

\begin{figure}
	\centering
	\subfloat{\includegraphics[width=0.5\textwidth]{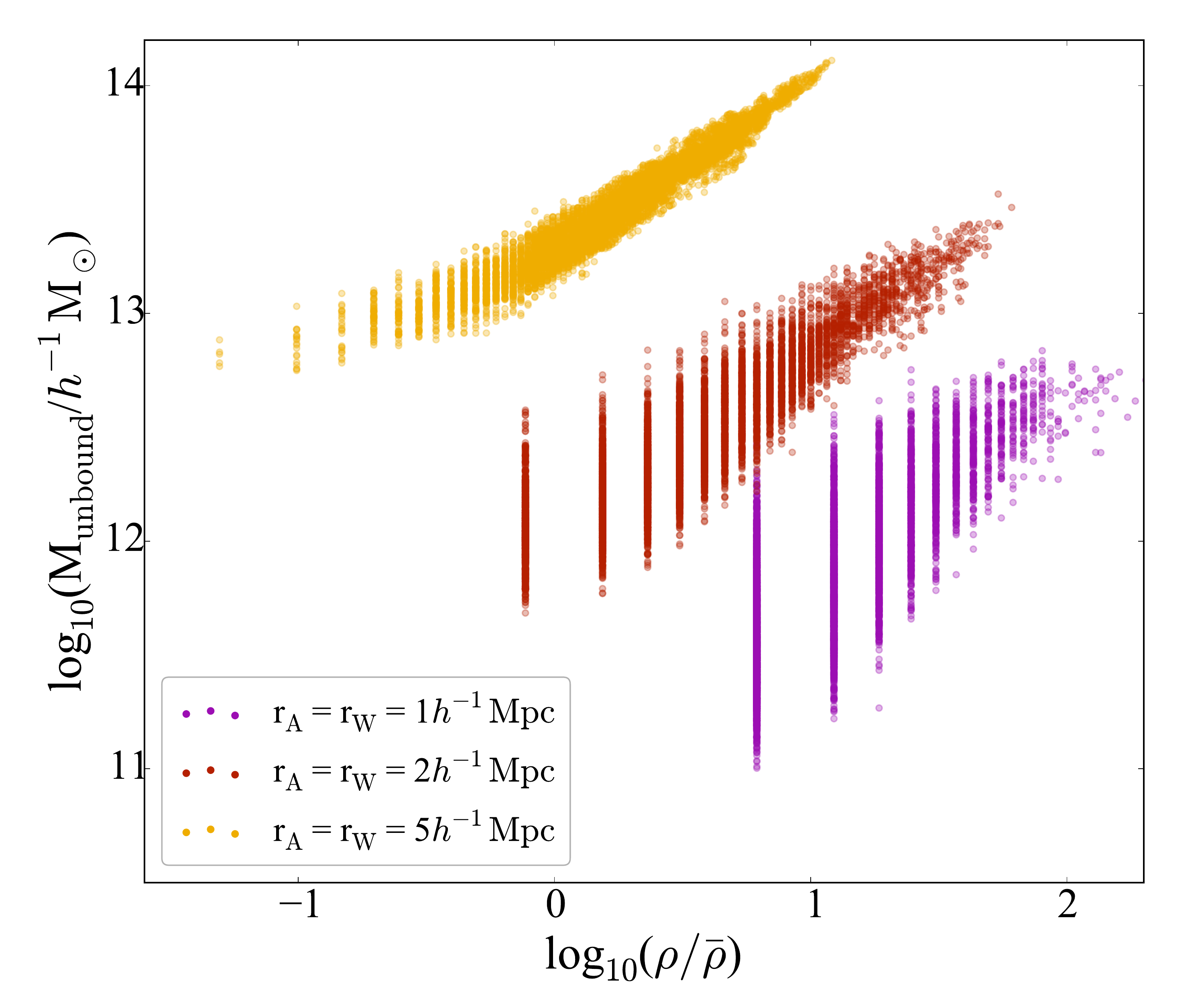}} \\
	\caption{The local density contrast, as measured by the number of galaxies within fixed apertures of 1 (purple), 2 (red), and 5 (yellow) \Mpc, versus the mass of the unbound matter contained within windows of the same radius at $z = 0$. \label{fig:Mub_FA}}
\end{figure}

Figure \ref{fig:Mub_FA} shows the fixed aperture galaxy count of the galaxies against the amount of unbound matter within apertures of \ra\ = 1, 2, and 5 \Mpc, shown by the purple, red, and yellow dots, respectively. Galaxy-count environment is by nature a quantised measure and we do not make allowances for part of a galaxy being within an aperture, resulting in discrete values for the density contrast at low density. These two quantities are closely correlated on all three scales. The largest radius aperture/window has the tightest correlation and the smallest radius has the weakest.

\begin{figure*}
	\centering
	\subfloat{\includegraphics[width=0.8\textwidth]{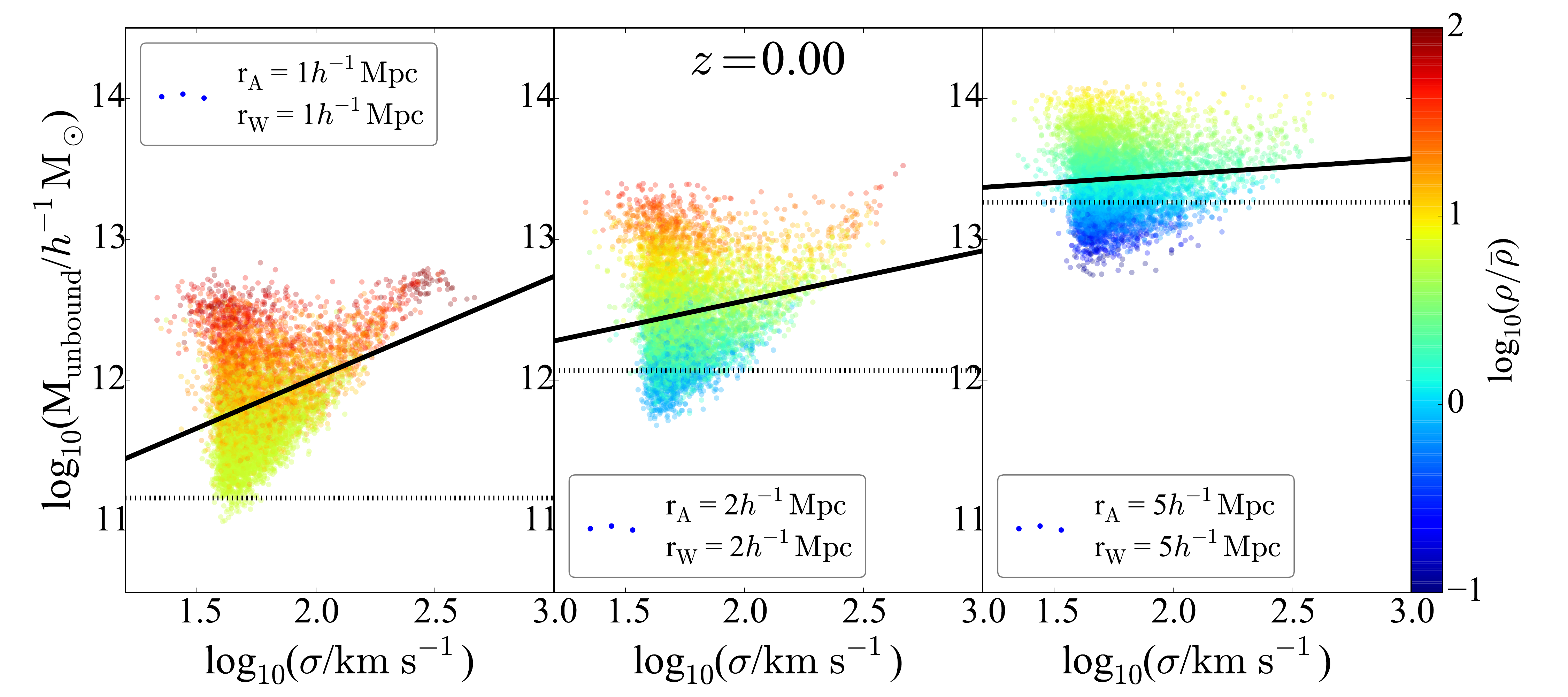}} \\
	\caption{Velocity dispersion versus the mass of the unbound matter within fixed windows of 1 (left), 2 (middle), and 3 (right) \Mpc\ at $z = 0$. These are the same distributions as the bottom row in Figure \ref{fig:Mub_fixed}, with velocity dispersion rather than \Mvir. The colour is the local density contrast, as measured by the number of galaxies within a fixed aperture of \ra\ = 1 (left), 2 (middle), and 3 (right) \Mpc. The solid black line is the fit to the data and the dotted black line is the expected mass of the unbound matter if it were evenly distributed. \label{fig:Mub_Vdisp_fixed_FA}}
\end{figure*}

To put the \env\ versus \Mub\ into the same context as our previous figures, we show in Figure \ref{fig:Mub_Vdisp_fixed_FA} the velocity dispersion of the halo against the mass of unbound matter within windows of \rw\ = 1, 2, and 5 \Mpc\ (left to right) at $z = 0$. We shift from using the halo's mass to its velocity dispersion, $\sigma$, which is the 1-dimensional velocity dispersion of the halo particles. It is related to the velocity dispersion of galaxies in a group or cluster and is therefore a more observable property. The black line is the same fit to the data as in Figure \ref{fig:Mub_fixed}, with the substitution of \Vdisp\ for \Mvir, and the dotted line is the expected amount of unbound mass in that window. The colour is based on the log of the density contrast, measured in apertures the same size as the windows.

The amount of unbound matter depends strongly on the environment of the halo and only weakly, if at all, on the halo velocity dispersion. The correlation between the halo and the nearby unbound matter measured in a fixed window appears to result largely from the correlation between halo mass and environment. At $z = 2$ the same pattern emerges but with less scatter, similar to the difference in redshifts of Figure \ref{fig:Mub_fixed}. The halos also inhabit a smaller range of \env s at higher redshift, which is predicted from structure formation. 

\begin{figure*}
	\centering
	\subfloat{\includegraphics[width=0.8\textwidth]{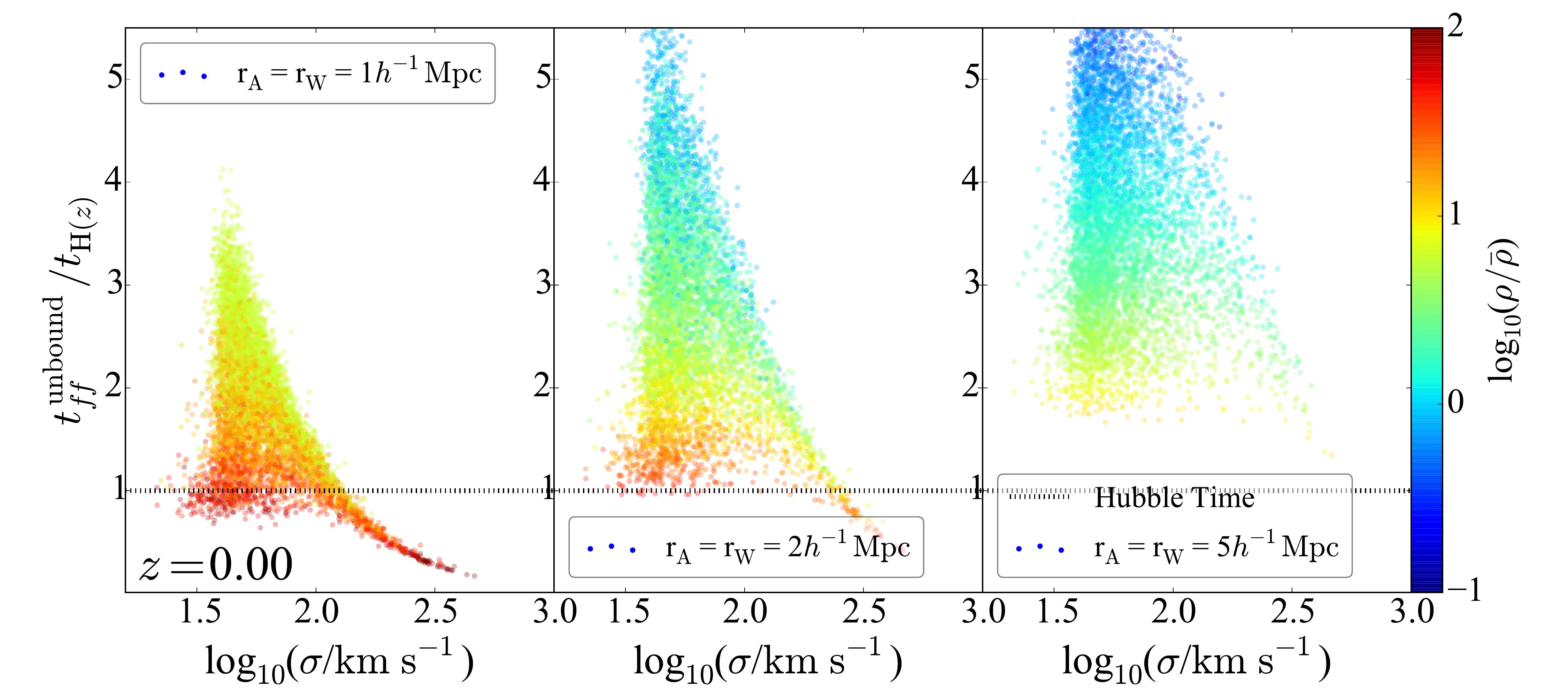}} \\
	\caption{Velocity dispersion versus the free fall time of the halo plus the unbound matter within fixed windows of 1 (left), 2 (middle), and 3 (right) \Mpc\ at $z = 0$. These are the same distributions as the bottom row in Figure \ref{fig:Mub_fixed_ff}, with velocity dispersion rather than \Mvir. The colour is the local density contrast, as measured by the number of galaxies within a fixed aperture of \ra\ = 1 (left), 2 (middle), and 3 (right) \Mpc. The dotted line is 1 Hubble time, or the age of the universe at $z = 0$. \label{fig:Mub_Vdisp_fixed_FA_ff}}
\end{figure*}

We see the same pattern in Figure \ref{fig:Mub_Vdisp_fixed_FA_ff}, which is the same as the bottom panel of Figure \ref{fig:Mub_fixed_ff}, with \Vdisp\ instead of \Mvir\ on the x-axis and is now coloured by the galaxy count \env\ as measured in apertures with radii the same size as the unbound matter windows, namely 1, 2, and 5 \Mpc\ (left to right, respectively). Here, the potential future evolution of the halo is highly dependent on the \env, with lower density halos (blue points) requiring far more time to accrete the nearby unbound matter as compared to halos in high density \env s (red points). This comes directly from the calculation of \tH\ in Equation \ref{eqn:free-fall-unbound}, where we showed the free fall time is dependent on both the amount of nearby unbound matter and the distance out to which the unbound matter extends.

\subsection{Stellar Mass and Other Galaxy Properties}
\label{unbound:mstars}
Thus far we have limited our measurements to the virial mass and velocity dispersion of a galaxy's halo, i.e. galaxy properties that emerge from the dark matter simulations alone. We now include observable properties from the baryonic physics in the semi-analytic model to test their interdependence on the amount of nearby unbound matter. We also use unbound matter as a proxy for the diffuse gas of the IGM, which can potentially be observed. To this end, we switch the y-axis from \Mub\ to $f_b \times $ \Mub\ to represent the diffuse gas, in a similar manner to \citet{Shattow2015}. 

The baryon fraction, $f_b$, is assumed to be the cosmic value of 0.18 \citep{Planck2015}. Semi-analytics show a baryon fraction for galaxies to be slightly higher than this (Croton et al., in prep), leaving the IGM a lower fraction overall. It is likely, however, that the deficit is found in regions of very low density, rather than in the volumes directly surrounding the halos. This results in the cosmic baryon fraction being a good estimate in the regions of unbound matter around halos.

\begin{figure}
	\centering
	\begin{tabular}{c}
		\vspace{-2mm}		\subfloat{\includegraphics[width=0.45\textwidth]{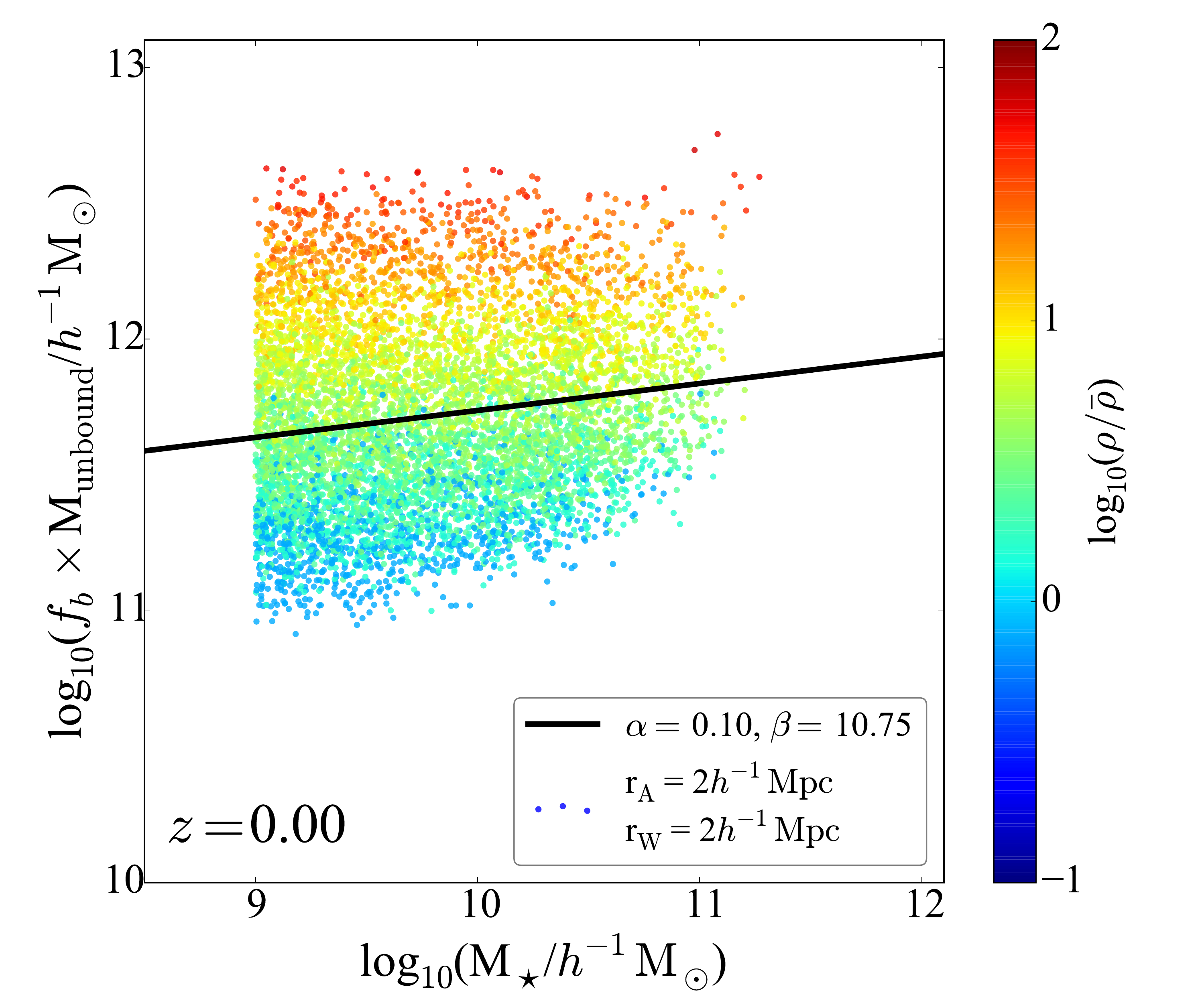}\label{fig:Mstar_FA}} \\
		\vspace{-2mm}
		\subfloat{\includegraphics[width=0.45\textwidth]{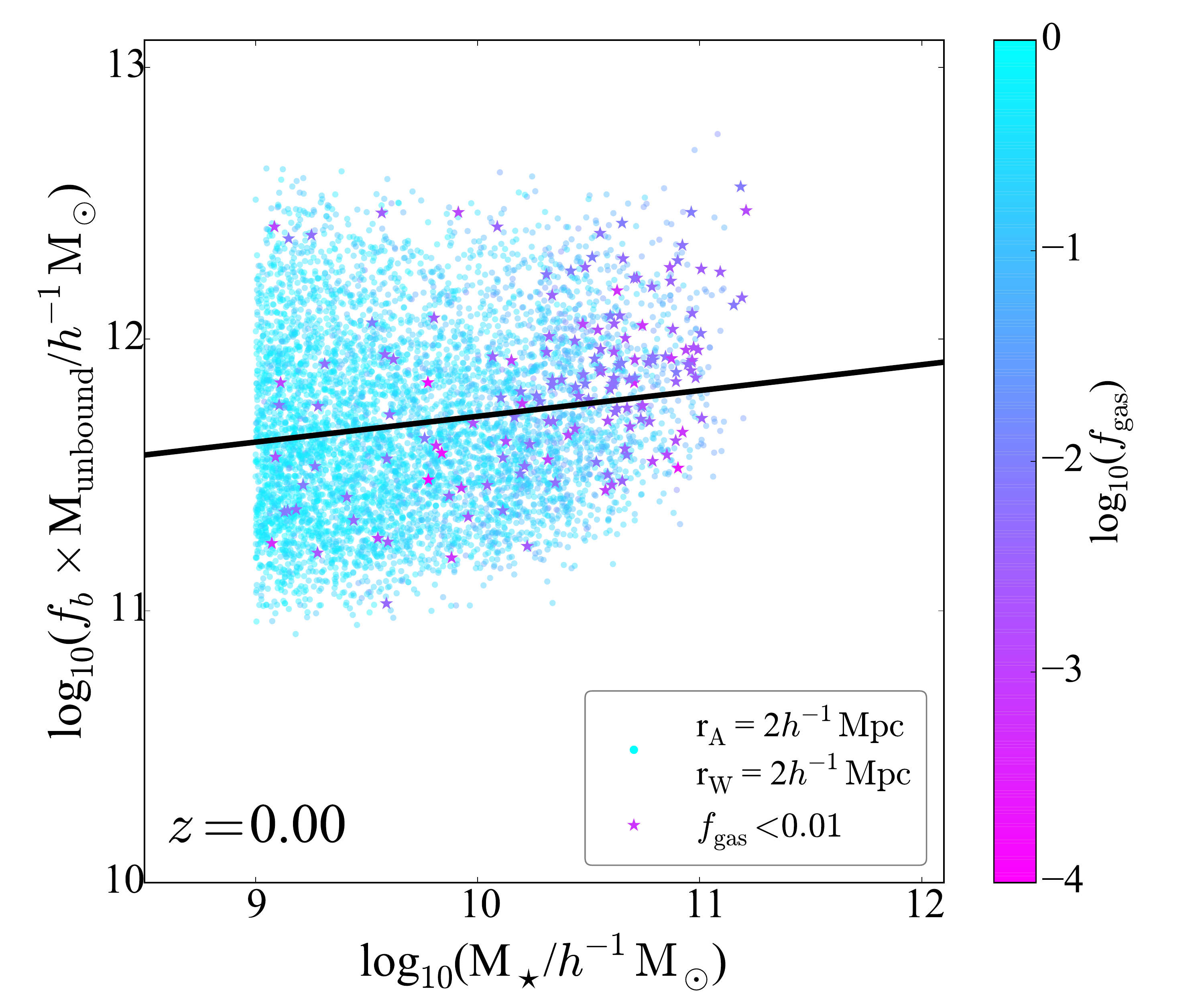}\label{fig:Mstar_fg}} \\
		\vspace{-2mm}		\subfloat{\includegraphics[width=0.45\textwidth]{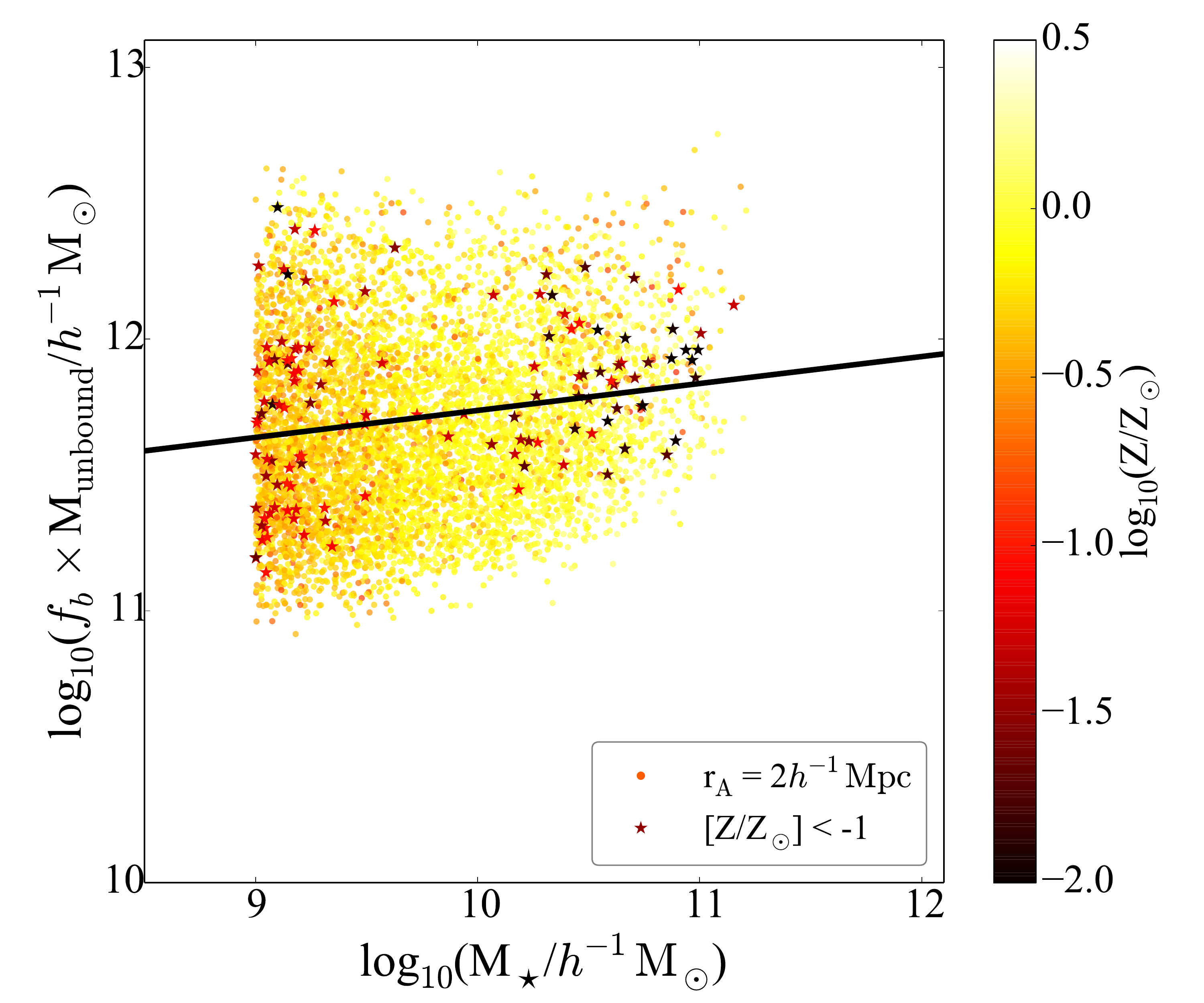}\label{fig:Mstar_ZZ}} \\
	\end{tabular}
	\caption[Unbound matter as a function of various observable galaxy properties]{Stellar mass versus the mass of the unbound baryonic matter found in fixed windows of radius \rw\ = 2 \Mpc\ at $z = 0$. Top panel: colours are the local density contrast measured by the number of galaxies within a 2 \Mpc\ fixed aperture. Middle panel: colours are the galaxy gas fraction. Galaxies with $f_{gas} < 0.01$ are tagged with stars. Bottom panel: colours are the metallicity of the galaxy disk. Galaxies with [Z/\Zsun] $< -1$ are marked with stars. \label{fig:Mub_Mstar}}
\end{figure}

Figure \ref{fig:Mub_Mstar} shows the stellar mass of a galaxy against the amount of diffuse gas (i.e. the unbound matter multiplied by the baryon fraction), within a 2 \Mpc\ window. The black lines are the fit to the data, with the slope and intercept listed in the top panel. All three panels are identical except for the colour coding, which represents a different observable in each panel: (top) the galaxy-count environment measured in a 2 \Mpc\ aperture, (middle) the gas fraction of the galaxy, and (bottom) the metallicity of the galaxy disk, all at $z = 0$.

Considering first all three panels together, the diffuse gas mass has a lower limit which increases slightly with stellar mass, as is the case in Figure \ref{fig:Mub_fixed}, although it is not as striking as the halo mass dependent limit there. The slope of the distribution is 0.10, $\sim 30\%$ less than the slope for the comparison against \Mvir, using the same window. The scatter is also significantly larger with \Mstar. Stellar mass, unlike \env, is therefore a poor predictor of unbound matter within a fixed window.

The top panel, colour-coded to show galaxy-count \env, reflects the findings of Figures \ref{fig:Mub_FA} and \ref{fig:Mub_Vdisp_fixed_FA} above, namely that galaxy-count environment closely tracks the amount of unbound matter, regardless of galaxy mass.  In other words, for a fixed stellar mass, the \env\ and the amount of nearby diffuse gas are strongly correlated.

The middle is colour-coded to show the gas fraction within the galaxy, where $f_{gas}$ = M$_{gas}$/(M$_{\star}$ + M$_{gas}$). We have tagged the galaxies with gas fractions of less than 1\% with pink stars. There is no discernible trend in these galaxies. Similar figures tagging the high gas fraction galaxies have shown the same result.

The bottom panel is colour-coded to show the metallicity of the cold gas in the disk of each galaxy. Metallicity is defined as Z = M$_\mathrm{Z}$/M$_\mathrm{ColdGas}$\ and is in units of solar metallicity, \Zsun = 0.02 \citep{Grevesse1998}. Low metallicity galaxies are tagged with stars and show no trend with \Mub. The excess of low metallicity galaxies at low masses is due to the fact that new galaxies in the model form out of pristine gas and have not had much time to create metals. Like in the middle panel, there is no dependence on the enclosed diffuse gas mass.

\section{Discussion}
\label{unbound:discussion}

\subsection{Unbound Matter}
\label{unbound:disc_ub}
At the beginning of this paper, we asked three questions regarding the unbound matter in the universe. The first, regarding its distribution and evolution over time, is answered in Figures \ref{fig:halo_evolution} - \ref{fig:Mub_fixed}. When measured in a halo dependent window, the relation between the virial mass of a halo and the amount of unbound matter nearby becomes steeper at low redshift, indicating that unbound matter has clustered around the more massive halos at a faster rate than around less massive halos, confirming previous studies. 

The \Mvir-\Mub\ relation is also slightly steeper for smaller windows, i.e. \rw\ = 2 \Rvir, than it is for larger windows, i.e. \rw\ = 5 \Rvir, with the difference being more pronounced at $z = 0$ than at higher redshift. This narrowing of the distribution mimics the growth of bound structure, with the halo filling the nearby volume with material pulled in from the outer volume. There is a slight increase in scatter in the relations, both with increasing radius and with time, with the largest amount of scatter being measured at 5 \Rvir\ from the halo at $z = 0$.

The dark matter directly outside of the halo is controlled by two opposing forces - the gravity pulling it towards the halo, and the galaxy feedback processes pushing it away. Since our model is built on an N-body simulation and does not include feedback processes until we add the semi-analytic model, we are limited to analyzing the effects of gravity alone. Most of the literature on the interplay between baryons and halos focuses on the inner part of the halos, where baryons have the strongest effect and their influence is more easily observed. \cite{Duffy2010}, however, show the outer edges of galaxy sized halos are slightly affected by feedback, with weak supernova feedback decreasing the amount of dark matter at the edge and strong feedback from AGN or supernovae increasing it. It is unlikely that these feedback processes would push the unbound matter more than an additional virial radius beyond the halo, even in a small halo. The amount of unbound matter, including both ejected and not-yet-bound material, would therefore increase in the annulus between 1 and 2 \Rvir, but not affect the rest of the distribution by much.

Unlike in the halo dependent window, the slope of the relation between unbound and virial mass does not change much with redshift for a given fixed-size window. The scatter, however, does increase considerably. Both the scatter and the slope of this relation decrease with increasing radius, as the volume probed by the fixed window approaches the mean density. The scatter is especially prevalent around smaller mass halos, reflecting the variety of environments that low mass halos inhabit. For example, the group sized halo ($\sim 10^{13}$ \Msun) shown in Figure \ref{fig:halo_evolution} has several dwarf galaxy halos and subhalos within a few virial radii. This large halo does not live within the spheres of influence of the low mass halos, but it does live within a few Mpc. These dwarf galaxy sized halos, therefore, will have a similar amount of unbound matter around them as the group sized halo.

The virial radius is defined as the point at which a density fluctuation is 200 times the critical density. This is a somewhat arbitrary definition \cite[see][and references therein for a full discussion]{Shull2014a}, since the slope of the mass profile continues outside of the determined virial limit, as shown in Figure \ref{fig:halo_evolution}. If the virial radius were determined to be smaller, as \cite{Shull2014a} finds for the Milky Way, the amount of matter that would be considered unbound in these simulations would increase. If \Rvir\ were defined as the point where a halo reaches $180 \times \rho_{crit}$, another common definition, its value would be larger and the amount of unbound matter would be lower. Likewise, if the boundary of 200 times the mean mass density ($\Omega_m$) is used, the amount of unbound matter decreases even further, as discussed by \citet{vanDaalen2015}. Also, as discussed in Section \ref{unbound:sim}, the amount of unbound matter changes depending on which halo finder is used. These profiles and relations are therefore dependent on several different factors, which are more likely to affect the results of the halo based window than the fixed radius window.

\subsection{Free-fall Times}
\label{unbound:disc_ff}
The second question we ask, regarding the likelihood of a halo to accrete all of the unbound matter within a window, we answer using the free-fall times, as shown in Figures \ref{fig:Mub_radial_ff} and \ref{fig:Mub_fixed_ff}. The free-fall time is calculated for the matter in the halo plus the unbound matter within a certain window. Lower density regions have larger free-fall times, and the farther from the halo a window reaches, the lower the total density will be. This accounts for the increase of free-fall times for larger windows. The increase in scatter with window size and decreasing redshift is magnified in the free-fall times.

Other effects to consider include the fact that halos have grown between $z = 2$ and $z = 0$, so the halo dependent window around each halo covers more volume. To counter this, we are using comoving coordinates, which decrease the extent of the radius with redshift by a factor of $(1 + z)^{-1}$ relative to the physical extent, as shown in Table \ref{tab:Rvir_z}. The solid black circle representing 1 \Rvir\ in Figure \ref{fig:halo_evolution} shows the growth of the halo is the dominant effect of these two between $z = 2$ and $z = 0.5$, although the radius decreases again between $z = 0.5$ and $z = 0$.

We have marked the Hubble time on Figures \ref{fig:Mub_radial_ff} and \ref{fig:Mub_fixed_ff}, showing what fraction of halos will be able to accrete all of the unbound matter within a given window in a Hubble time, as measured at that redshift. It is also interesting to consider how many of the halos at $z = 2$ could have accreted all of their enclosed unbound matter by $z = 0$. $t_{\mathrm{H}}(z = 0) \simeq 3 \times t_{\mathrm{H}}(z = 2)$, so the unbound matter would have to be within 2 additional Hubble times (as measured at $z = 2$) for the halo to be able to accrete all of the enclosed matter. From these figures, we can see that the free fall times of the enclosed masses out to both 5 \Rvir\ and 5 \Mpc\ are less than 1 \tH\ at $z = 2$, well within the cutoff for accretion.

As mentioned above, the free-fall time measures the pressure-less collapse of a spherical distribution. Since we do not have a pressure-less system and it is obvious from Figure \ref{fig:halo_evolution} that we do not have a spherical object, it is interesting to note how many of these particles are actually accreted by halos between $z = 2$ and $z = 0$. 

By comparing the particles that are within 5 \Rvir\ of a halo at $z = 2$ to the particles that are part of a halo at $z = 0$, we find that 78\% of nearby particles are accreted. The same comparison to particles within 5 \Mpc\ at $z = 2$ shows that only 36\% of these proximal particles are accreted in the intervening time. The lower accretion rate using the fixed window size most likely results from the large number of multiply counted unbound particles in that metric, which increases the free-fall times when the particles are counted only once (see Section \ref{unbound:freefall}). Future work will take this into account for a more rigorous calculation of the free-fall times for the purpose of predicting future evolution of a halo.


If we assume baryons trace the dark matter particles as halos accrete them, then the very long free-fall times (which are likely lower limits on the accretion time-scales) that have emerged by $z = 0$ could indicate a natural quenching of the growth of the interior galaxy that has nothing to do with baryonic processes.  Halos are merely accreting material more slowly, which will increasingly slow down the star formation cycle on which the galaxy depends. In this case, semi-analytic models might assume too high an efficiency for feedback processes when, in reality, halos might quench themselves. This idea is partially backed up by Figure \ref{fig:Mub_Vdisp_fixed_FA_ff}, which shows that the local galaxy count density might play a large role in the future evolution of the galaxy.

\subsection{Observables}
\label{unbound:disc_obs}
Our third question addresses 
if there are any correlations between observable properties of a galaxy and the amount of unbound matter, with the intention of using these correlations to make predictions for the future observation of the unbound structure of the Universe.

We first look at the unbound matter against the galaxy count \env, both on its own and as a function of the velocity dispersion of the halo. These two properties are closely related, with the strongest correlation at a scale of \rw\ = \ra\ = 5 \Mpc. Thus, a 5 \Mpc\ aperture is a good means of determining how much unbound matter is inside a window of the same size, regardless of galaxy mass. Taking Figure \ref{fig:Mub_fixed} into account, however, shows that this scale probes very close to the mean density of unbound matter. Its usefulness as a probe of future growth is therefore less certain.

From Figure \ref{fig:Mub_fixed_ff}, we see that zero halos at $z = 0$ have the potential to accrete all of the unbound matter out to 5 \Mpc\ by the time the universe doubles in age, i.e. within another Hubble time as measured at $z = 0$. Even on a 1 \Mpc\ scale, only 12\% of the galaxies can accrete all of the unbound matter on this timescale, making the smaller radius a more physical choice of window/aperture size to follow to determine future growth, although having a weaker correlation and more scatter means it is less clear how much unbound matter surrounds the halo based on its galaxy-count \env.

Interestingly, halos of all masses with more nearby galaxies also have more unbound matter, and therefore shorter free fall times. Rather than depleting the unbound matter by accreting it all onto bound halos, for a halo of a given mass having more nearby galaxies is a strong indicator of relatively more unbound mass within the window. This is especially apparent in Figures \ref{fig:Mub_FA} - \ref{fig:Mub_Vdisp_fixed_FA_ff}.

This effect is even more pronounced when galaxy stellar mass is compared to the diffuse IGM in Figure \ref{fig:Mub_Mstar}. There is little correlation between stellar mass and diffuse gas in a fixed radius window, since galaxies of a given mass can occupy a range of halo masses, even as the central galaxy. Any pattern in the halo mass and the unbound dark matter from Figure \ref{fig:Mub_fixed} is therefore washed out. The galaxy count environment, however, shows a clear trend with unbound matter in the top panel of \ref{fig:Mub_Mstar}. Of all of the possible observable properties we can measure using the model, the galaxy count environment is the best indicator of how much unbound matter is within a window of the same size.

Different calculations of \env\ often correlate with one another \citep[e.g.][]{Muldrew2012,Shattow2013} because their goal is to probe the underlying structure of the mass density field. Most environment metrics are limited to observables, such as the number of galaxies nearby, the distance to the Nth nearest neighbour, or the amount of volume closer to a galaxy than to any other galaxy (i.e. Voronoi tessellation). These metrics all use galaxies, or the peaks of the density field as their density defining populations. This work shows that the relation between environment metrics continues to lower density regions of space - where the diffuse gas lives.

Gas fraction and metallicity are also associated with stellar mass, and to a lesser extent, \env\ \citep{Thomas2005, DeLucia2006, Cooper2008b, Dave2011}. These two galaxy properties, however, show no correlation with nearby unbound matter beyond their initial dependence on stellar mass. Since they do correlate with \env, one might expect some variation with unbound matter. The lack of a trend is possibly due to the type of galaxy model used. Semi-analytic models do very well at reproducing observable properties and relations of galaxies, but they are based entirely on the halos. They therefore lack environment information beyond the virial radius of the halo in which the galaxy resides. A more complex model or a hydrodynamic simulation might be required to tease out these effects.

Overall, galaxy count environment is the closest measurement we have to a smoking gun of nearby unbound matter, especially when they are both measured on the same scale.

\section{Summary}
\label{unbound:summary}

This paper addresses the often-ignored unbound matter in simulations in the context of how a galaxy's (or halo's) \env\ is affected by its evolution. We investigate the relationship between the amount of unbound matter and halo mass using a variable and a fixed window and examine how these relationships evolve over time. We also consider at the potential of a halo to accrete the nearby unbound matter within a Hubble time. Finally, we compare the unbound matter to observable properties of galaxies in the hopes of finding a clear signal of nearby unbound matter for observers to look for.

The main conclusions of this paper are:

\begin{itemize}
	\item Unbound matter clusters around bound matter, with more unbound matter being found near massive halos when measured on a fixed scale, as seen in Figure \ref{fig:Mub_fixed}. For halo dependent windows, this is only true from \rw\ = 2 \Rvir\ at $z = 0$. The remaining windows at both redshifts have slopes equal to or less than 1, meaning there is no preference for more massive halos. This is shown in Figure \ref{fig:Mub_radial}.
	\item When measured in windows scaled to the virial radius of the halo, the free-fall time of the total mass inside the window reveals that most halos have the potential to accrete all of their surrounding unbound matter within a Hubble time out to a radius of 5 \Rvir. We show this in Figure \ref{fig:Mub_radial_ff}.
	\item At $z = 2$, all galaxies can accrete all of the diffuse matter out to a fixed radius of 5 \Mpc\ within a Hubble time. At $z = 0$, only 12\% can accrete all of the unbound matter out to 1 \Mpc, as seen in Figure \ref{fig:Mub_fixed_ff}.
	\item Galaxies in denser environments are more likely to have more unbound mass nearby, regardless of stellar or halo mass, which we show in Figures \ref{fig:Mub_Vdisp_fixed_FA}, \ref{fig:Mub_Vdisp_fixed_FA_ff}, and \ref{fig:Mub_Mstar}.
	\item Other galaxy properties, such as gas fraction and disk metallicity, show little correlation with the diffuse gas, making galaxy-count \env\ the most promising observational probe of the large scale distribution of the intergalactic medium in our analysis, as is clear in Figure \ref{fig:Mub_Mstar}.
\end{itemize}

Future work will look at the evolution histories of specific halos, measuring how important these accretion events are in comparison to other means of growing, namely major and minor mergers. This will further address the question of how a galaxy's evolution is affected by its \env.

\section*{Acknowledgments}
The authors would like to thank Antonio Bibiano and the anonymous referee for their insightful comments.

GMS is supported by a Swinburne University SUPRA postgraduate scholarship. DC acknowledges the receipt of a QEII Fellowship from the Australian Research Council.

The Millennium and milli-Millennium Simulations used in this paper were carried out by the Virgo Supercomputing Consortium at the Computing Centre of the Max Planck Society. The databases used in this paper and the web application providing online access to them were constructed as part of the activities of the German Astrophysical Virtual Observatory (GAVO). 

\bibliographystyle{mn2e}
\bibliography{refs}

\begin{thebibliography}{44}
\expandafter\ifx\csname natexlab\endcsname\relax\def\natexlab#1{#1}\fi

\bibitem[{{Angulo} \& {White}(2010)}]{Angulo2010}
{Angulo} R.~E., {White} S.~D.~M., 2010, \mnras, 401, 1796

\bibitem[{{Behroozi} {et~al}\mbox{.}(2013){Behroozi}, {Wechsler}, \&
  {Wu}}]{Behroozi2013}
{Behroozi} P.~S., {Wechsler} R.~H., {Wu} H.-Y., 2013, \apj, 762, 109

\bibitem[{{Birnboim} \& {Dekel}(2003)}]{Birnboim2003}
{Birnboim} Y., {Dekel} A., 2003, \mnras, 345, 349

\bibitem[{{Cantalupo} {et~al}\mbox{.}(2014){Cantalupo}, {Arrigoni-Battaia},
  {Prochaska}, {Hennawi}, \& {Madau}}]{Cantalupo2014}
{Cantalupo} S., {Arrigoni-Battaia} F., {Prochaska} J.~X., {Hennawi} J.~F.,
  {Madau} P., 2014, \nat, 506, 63

\bibitem[{{Cooper} {et~al}\mbox{.}(2008){Cooper}, {Tremonti}, {Newman}, \&
  {Zabludoff}}]{Cooper2008b}
{Cooper} M.~C., {Tremonti} C.~A., {Newman} J.~A., {Zabludoff} A.~I., 2008,
  \mnras, 390, 245

\bibitem[{{Crain} {et~al}\mbox{.}(2009){Crain}, {Theuns}, {Dalla Vecchia},
  {Eke}, {Frenk}, {Jenkins}, {Kay}, {Peacock}, {Pearce}, {Schaye}, {Springel},
  {Thomas}, {White}, \& {Wiersma}}]{Crain2009}
{Crain} R.~A. {et~al.}, 2009, \mnras, 399, 1773

\bibitem[{{Croton} {et~al}\mbox{.}(2006){Croton}, {Springel}, {White}, {De
  Lucia}, {Frenk}, {Gao}, {Jenkins}, {Kauffmann}, {Navarro}, \&
  {Yoshida}}]{Croton2006}
{Croton} D.~J. {et~al.}, 2006, \mnras, 365, 11

\bibitem[{{Dav{\'e}} {et~al}\mbox{.}(2011){Dav{\'e}}, {Finlator}, \&
  {Oppenheimer}}]{Dave2011}
{Dav{\'e}} R., {Finlator} K., {Oppenheimer} B.~D., 2011, \mnras, 416, 1354

\bibitem[{{Dav{\'e}} {et~al}\mbox{.}(1999){Dav{\'e}}, {Hernquist}, {Katz}, \&
  {Weinberg}}]{Dave1999}
{Dav{\'e}} R., {Hernquist} L., {Katz} N., {Weinberg} D.~H., 1999, \apj, 511,
  521

\bibitem[{{Dav{\'e}} {et~al}\mbox{.}(2010){Dav{\'e}}, {Oppenheimer}, {Katz},
  {Kollmeier}, \& {Weinberg}}]{Dave2010}
{Dav{\'e}} R., {Oppenheimer} B.~D., {Katz} N., {Kollmeier} J.~A., {Weinberg}
  D.~H., 2010, \mnras, 408, 2051

\bibitem[{{De Lucia} {et~al}\mbox{.}(2006){De Lucia}, {Springel}, {White},
  {Croton}, \& {Kauffmann}}]{DeLucia2006}
{De Lucia} G., {Springel} V., {White} S.~D.~M., {Croton} D., {Kauffmann} G.,
  2006, \mnras, 366, 499

\bibitem[{{Duffy} {et~al}\mbox{.}(2010){Duffy}, {Schaye}, {Kay}, {Dalla
  Vecchia}, {Battye}, \& {Booth}}]{Duffy2010}
{Duffy} A.~R., {Schaye} J., {Kay} S.~T., {Dalla Vecchia} C., {Battye} R.~A.,
  {Booth} C.~M., 2010, \mnras, 405, 2161

\bibitem[{{Fakhouri} \& {Ma}(2008)}]{Fakhouri2008}
{Fakhouri} O., {Ma} C.-P., 2008, \mnras, 386, 577

\bibitem[{{Fakhouri} \& {Ma}(2010)}]{Fakhouri2010}
{Fakhouri} O., {Ma} C.-P., 2010, \mnras, 401, 2245

\bibitem[{{Genel} {et~al}\mbox{.}(2010){Genel}, {Bouch{\'e}}, {Naab},
  {Sternberg}, \& {Genzel}}]{Genel2010}
{Genel} S., {Bouch{\'e}} N., {Naab} T., {Sternberg} A., {Genzel} R., 2010,
  \apj, 719, 229

\bibitem[{{Grevesse} \& {Sauval}(1998)}]{Grevesse1998}
{Grevesse} N., {Sauval} A.~J., 1998, \ssr, 85, 161

\bibitem[{{Haas} {et~al}\mbox{.}(2012){Haas}, {Schaye}, \&
  {Jeeson-Daniel}}]{Haas2012}
{Haas} M.~R., {Schaye} J., {Jeeson-Daniel} A., 2012, \mnras, 419, 2133

\bibitem[{{Kere{\v s}} {et~al}\mbox{.}(2005){Kere{\v s}}, {Katz}, {Weinberg},
  \& {Dav{\'e}}}]{Keres2005}
{Kere{\v s}} D., {Katz} N., {Weinberg} D.~H., {Dav{\'e}} R., 2005, \mnras, 363,
  2

\bibitem[{{Knollmann} \& {Knebe}(2009)}]{Knollmann2009}
{Knollmann} S.~R., {Knebe} A., 2009, \apjs, 182, 608

\bibitem[{{Lacey} \& {Cole}(1993)}]{Lacey1993}
{Lacey} C., {Cole} S., 1993, \mnras, 262, 627

\bibitem[{{Lemson} \& {Kauffmann}(1999)}]{Lemson1999}
{Lemson} G., {Kauffmann} G., 1999, \mnras, 302, 111

\bibitem[{{Lotz} {et~al}\mbox{.}(2011){Lotz}, {Jonsson}, {Cox}, {Croton},
  {Primack}, {Somerville}, \& {Stewart}}]{Lotz2011}
{Lotz} J.~M., {Jonsson} P., {Cox} T.~J., {Croton} D., {Primack} J.~R.,
  {Somerville} R.~S., {Stewart} K., 2011, \apj, 742, 103

\bibitem[{{Lynds}(1971)}]{Lynds1971}
{Lynds} R., 1971, \apjl, 164, L73

\bibitem[{Ma {et~al}\mbox{.}(2009)Ma, Ebeling, \& Barrett}]{Ma2009}
Ma C.-J., Ebeling H., Barrett E., 2009, The Astrophysical Journal Letters, 693,
  L56

\bibitem[{{Macci{\`o}} {et~al}\mbox{.}(2007){Macci{\`o}}, {Dutton}, {van den
  Bosch}, {Moore}, {Potter}, \& {Stadel}}]{Maccio2007}
{Macci{\`o}} A.~V., {Dutton} A.~A., {van den Bosch} F.~C., {Moore} B., {Potter}
  D., {Stadel} J., 2007, \mnras, 378, 55

\bibitem[{Martin {et~al}\mbox{.}(2014)Martin, Chang, Matuszewski, Morrissey,
  Rahman, Moore, \& Steidel}]{Martin2014}
Martin D.~C., Chang D., Matuszewski M., Morrissey P., Rahman S., Moore A.,
  Steidel C.~C., 2014, The Astrophysical Journal, 786, 106

\bibitem[{{Mihos} \& {Hernquist}(1994)}]{Mihos1994}
{Mihos} J.~C., {Hernquist} L., 1994, \apjl, 425, L13

\bibitem[{{Muldrew} {et~al}\mbox{.}(2012){Muldrew}, {Croton}, {Skibba},
  {Pearce}, {Ann}, {Baldry}, {Brough}, {Choi}, {Conselice}, {Cowan},
  {Gallazzi}, {Gray}, {Gr{\"u}tzbauch}, {Li}, {Park}, {Pilipenko}, {Podgorzec},
  {Robotham}, {Wilman}, {Yang}, {Zhang}, \& {Zibetti}}]{Muldrew2012}
{Muldrew} S.~I. {et~al.}, 2012, \mnras, 419, 2670

\bibitem[{{Planck Collaboration} {et~al}\mbox{.}(2013){Planck Collaboration},
  {Ade}, {Aghanim}, {Arnaud}, {Ashdown}, {Atrio-Barandela}, {Aumont},
  {Baccigalupi}, {Balbi}, {Banday}, \& et~al.}]{Planck2013a}
{Planck Collaboration} {et~al.}, 2013, \aap, 550, A134

\bibitem[{{Planck Collaboration} {et~al}\mbox{.}(2015){Planck Collaboration},
  {Ade}, {Aghanim}, {Arnaud}, {Ashdown}, {Aumont}, {Baccigalupi}, {Banday},
  {Barreiro}, {Bartlett}, \& et~al.}]{Planck2015}
{Planck Collaboration} {et~al.}, 2015, ArXiv e-prints

\bibitem[{{Press} \& {Schechter}(1974)}]{Press1974}
{Press} W.~H., {Schechter} P., 1974, \apj, 187, 425

\bibitem[{{Shattow} {et~al}\mbox{.}(2015){Shattow}, {Croton}, \&
  {Bibiano}}]{Shattow2015}
{Shattow} G.~M., {Croton} D.~J., {Bibiano} A., 2015, \mnras, 450, 2306

\bibitem[{{Shattow} {et~al}\mbox{.}(2013){Shattow}, {Croton}, {Skibba},
  {Muldrew}, {Pearce}, \& {Abbas}}]{Shattow2013}
{Shattow} G.~M., {Croton} D.~J., {Skibba} R.~A., {Muldrew} S.~I., {Pearce}
  F.~R., {Abbas} U., 2013, \mnras, 433, 3314

\bibitem[{{Shull}(2014)}]{Shull2014a}
{Shull} J.~M., 2014, \apj, 784, 142

\bibitem[{{Spergel} {et~al}\mbox{.}(2003){Spergel}, {Verde}, {Peiris},
  {Komatsu}, {Nolta}, {Bennett}, {Halpern}, {Hinshaw}, {Jarosik}, {Kogut},
  {Limon}, {Meyer}, {Page}, {Tucker}, {Weiland}, {Wollack}, \&
  {Wright}}]{Spergel2003}
{Spergel} D.~N. {et~al.}, 2003, \apjs, 148, 175

\bibitem[{{Springel} {et~al}\mbox{.}(2005){Springel}, {White}, {Jenkins},
  {Frenk}, {Yoshida}, {Gao}, {Navarro}, {Thacker}, {Croton}, {Helly},
  {Peacock}, {Cole}, {Thomas}, {Couchman}, {Evrard}, {Colberg}, \&
  {Pearce}}]{Springel2005a}
{Springel} V. {et~al.}, 2005, \nat, 435, 629

\bibitem[{{Springel} {et~al}\mbox{.}(2001){Springel}, {White}, {Tormen}, \&
  {Kauffmann}}]{Springel2001}
{Springel} V., {White} S.~D.~M., {Tormen} G., {Kauffmann} G., 2001, \mnras,
  328, 726

\bibitem[{Takeuchi {et~al}\mbox{.}(2014)Takeuchi, Zaroubi, \&
  Sugiyama}]{Takeuchi2014}
Takeuchi Y., Zaroubi S., Sugiyama N., 2014, Monthly Notices of the Royal
  Astronomical Society, 444, 2236

\bibitem[{{Thomas} {et~al}\mbox{.}(2005){Thomas}, {Maraston}, {Bender}, \&
  {Mendes de Oliveira}}]{Thomas2005}
{Thomas} D., {Maraston} C., {Bender} R., {Mendes de Oliveira} C., 2005, \apj,
  621, 673

\bibitem[{{Tremonti} {et~al}\mbox{.}(2004){Tremonti}, {Heckman}, {Kauffmann},
  {Brinchmann}, {Charlot}, {White}, {Seibert}, {Peng}, {Schlegel}, {Uomoto},
  {Fukugita}, \& {Brinkmann}}]{Tremonti2004}
{Tremonti} C.~A. {et~al.}, 2004, \apj, 613, 898

\bibitem[{{van Daalen} \& {Schaye}(2015)}]{vanDaalen2015}
{van Daalen} M.~P., {Schaye} J., 2015, ArXiv e-prints

\bibitem[{{Wang} {et~al}\mbox{.}(2011){Wang}, {Mo}, {Jing}, {Yang}, \&
  {Wang}}]{Wang2011}
{Wang} H., {Mo} H.~J., {Jing} Y.~P., {Yang} X., {Wang} Y., 2011, \mnras, 413,
  1973

\bibitem[{{Wechsler} {et~al}\mbox{.}(2002){Wechsler}, {Bullock}, {Primack},
  {Kravtsov}, \& {Dekel}}]{Wechsler2002}
{Wechsler} R.~H., {Bullock} J.~S., {Primack} J.~R., {Kravtsov} A.~V., {Dekel}
  A., 2002, \apj, 568, 52

\bibitem[{{Werner} {et~al}\mbox{.}(2008){Werner}, {Finoguenov}, {Kaastra},
  {Simionescu}, {Dietrich}, {Vink}, \& {B{\"o}hringer}}]{Werner2008}
{Werner} N., {Finoguenov} A., {Kaastra} J.~S., {Simionescu} A., {Dietrich}
  J.~P., {Vink} J., {B{\"o}hringer} H., 2008, \aap, 482, L29

\end{thebibliography}

\end{document}